\pdfoutput=1
\documentclass[english, aps, prb, reprint, superscriptaddress]{revtex4-1}
\usepackage[LGR,T1]{fontenc}
\usepackage[latin9]{inputenc}
\usepackage{color}
\usepackage{babel}
\usepackage{amstext}
\usepackage{graphicx}
\usepackage{subscript}
\usepackage[unicode=true,pdfusetitle,
 bookmarks=true,bookmarksnumbered=false,bookmarksopen=false,
 breaklinks=false,pdfborder={0 0 1},backref=false,colorlinks=true]
 {hyperref}
\hypersetup{
 allcolors=blue}

\makeatletter

\DeclareRobustCommand{\greektext}{%
  \fontencoding{LGR}\selectfont\def\encodingdefault{LGR}}
\DeclareRobustCommand{\textgreek}[1]{\leavevmode{\greektext #1}}
\ProvideTextCommand{\~}{LGR}[1]{\char126#1}

\usepackage[all]{hypcap}

\makeatother

\begin{document}

\title{Single-Shot Readout Performance of Two Heterojunction-Bipolar-Transistor
Amplification Circuits at Millikelvin Temperatures}

\author{M. J. Curry}

\thanks{M. J. Curry and M. Rudolph contributed equally to this work.}

\affiliation{Department of Physics and Astronomy, University of New Mexico, Albuquerque,
New Mexico, 87131, USA}

\affiliation{Center for Quantum Information and Control, University of New Mexico,
Albuquerque, New Mexico, 87131, USA}

\affiliation{Sandia National Laboratories, 1515 Eubank Blvd SE, Albuquerque, New
Mexico, 87123, USA}

\author{M. Rudolph}

\thanks{M. J. Curry and M. Rudolph contributed equally to this work.}

\affiliation{Sandia National Laboratories, 1515 Eubank Blvd SE, Albuquerque, New
Mexico, 87123, USA}

\author{T. D. England}

\affiliation{Sandia National Laboratories, 1515 Eubank Blvd SE, Albuquerque, New
Mexico, 87123, USA}

\author{A. M. Mounce}

\affiliation{Sandia National Laboratories, 1515 Eubank Blvd SE, Albuquerque, New
Mexico, 87123, USA}

\author{R. M. Jock}

\affiliation{Sandia National Laboratories, 1515 Eubank Blvd SE, Albuquerque, New
Mexico, 87123, USA}

\author{C. Bureau-Oxton}

\affiliation{Département de physique et Institut quantique, Université de Sherbrooke,
Sherbrooke (Québec) J1K 2R1, Canada}

\affiliation{Sandia National Laboratories, 1515 Eubank Blvd SE, Albuquerque, New
Mexico, 87123, USA}

\author{P. Harvey-Collard}

\affiliation{Département de physique et Institut quantique, Université de Sherbrooke,
Sherbrooke (Québec) J1K 2R1, Canada}

\affiliation{Sandia National Laboratories, 1515 Eubank Blvd SE, Albuquerque, New
Mexico, 87123, USA}

\author{P. A. Sharma}

\affiliation{Sandia National Laboratories, 1515 Eubank Blvd SE, Albuquerque, New
Mexico, 87123, USA}

\author{J. M. Anderson}

\affiliation{Sandia National Laboratories, 1515 Eubank Blvd SE, Albuquerque, New
Mexico, 87123, USA}

\author{D. M. Campbell}

\affiliation{Sandia National Laboratories, 1515 Eubank Blvd SE, Albuquerque, New
Mexico, 87123, USA}

\author{J. R. Wendt}

\affiliation{Sandia National Laboratories, 1515 Eubank Blvd SE, Albuquerque, New
Mexico, 87123, USA}

\author{D. R. Ward}

\affiliation{Sandia National Laboratories, 1515 Eubank Blvd SE, Albuquerque, New
Mexico, 87123, USA}

\author{S. M. Carr}

\affiliation{Sandia National Laboratories, 1515 Eubank Blvd SE, Albuquerque, New
Mexico, 87123, USA}

\author{M. P. Lilly}

\affiliation{Sandia National Laboratories, 1515 Eubank Blvd SE, Albuquerque, New
Mexico, 87123, USA}

\affiliation{Center for Integrated Nanotechnologies, 1515 Eubank Blvd SE, Albuquerque,
New Mexico, 87123, USA}

\author{M. S. Carroll}

\affiliation{Sandia National Laboratories, 1515 Eubank Blvd SE, Albuquerque, New
Mexico, 87123, USA}
\begin{abstract}
High-fidelity single-shot readout of spin qubits requires distinguishing
states much faster than the T\textsubscript{1} time of the spin state.
One approach to improving readout fidelity and bandwidth (BW) is cryogenic
amplification, where the signal from the qubit is amplified before
noise sources are introduced and room-temperature amplifiers can operate
at lower gain and higher BW. We compare the performance of two cryogenic
amplification circuits: a current-biased heterojunction bipolar transistor
circuit (CB-HBT), and an AC-coupled HBT circuit (AC-HBT). Both circuits
are mounted on the mixing-chamber stage of a dilution refrigerator
and are connected to silicon metal oxide semiconductor (Si-MOS) quantum
dot devices on a printed circuit board (PCB). The power dissipated
by the CB-HBT ranges from 0.1 to 1 $\mu$W whereas the power of the
AC-HBT ranges from 1 to 20 $\mu$W. Referred to the input, the noise
spectral density is low for both circuits, in the 15 to 30 fA/$\sqrt{\textrm{Hz}}$
range. The charge sensitivity for the CB-HBT and AC-HBT is 330 $\mu$e/$\sqrt{\textrm{Hz}}$
and 400 $\mu$e/$\sqrt{\textrm{Hz}}$, respectively. For the single-shot
readout performed, less than 10 $\mu$s is required for both circuits
to achieve bit error rates below 10\textsuperscript{-3}, which is
a putative threshold for quantum error correction.
\end{abstract}
\maketitle

\section*{Introduction}

Spin qubits in semiconductors are a promising platform for building
quantum computers \citep{Kane_1998,Elzerman_2004,Petta_2005,Morello_2010,Pla_2013,Eng_2015,Zajac_2017,Rochette_2017}.
Significant progress has been achieved in recent years, including
demonstrations of extremely long coherence times \citep{Muhonen_2014},
high-fidelity state readout \citep{Harvey-Collard_2018readout,Nakajima_2017,Shulman_2014,Watson_2015},
high-fidelity single qubits gates \citep{Takeda_2016,Muhonen_2014,Kawakami_2016,Nichol_2017},
and two qubit gates \citep{Shulman_2012,Nichol_2017,Yoneda_2018,Zajac_2017}.
As the field advances to multiple qubit systems, improvements in single-shot
state readout and measurement times will be necessary to achieve fault
tolerance. Improving the signal-to-noise ratio (SNR) and bandwidth
(BW) of the qubit state detection is critical for both tunnel rate
selective readout \citep{Elzerman_2004} and energy selective readout
\citep{Petta_2005}. With the same bit error rate, faster readout
will reduce tunnel rate and metastable relaxation or relaxation related
errors.

Cryogenic amplification is one way the readout SNR and BW can be improved.
Challenges are that: 1) input signals remain relatively small \citep{Onac_2006,Khrapai_2006,Gustavsson_2007,Horibe_2015,Harvey-Collard_2017qubit}
and 2) significant noise and parasitic capacitance is introduced into
the measurement circuit when routing the signal out of a dilution
refrigerator \citep{Kalra_2016}. Several approaches for cryogenic
amplification include: radio-frequency (RF) resonant quantum point
contact (QPC) and single electron transistor (SET) circuits \citep{Schoelkopf_1998,Aassime_2001,Reilly_2007,Barthel_2009,Barthel_2010,Mason_2010,Yuan_2012,Verduijn_2014},
gate dispersive RF circuits \citep{Colless_2013}, Josephson parametric
amplification circuits \citep{Stehlik_2015}, and cryogenic transistors
\citep{Visscher_1996,Pettersson_1996,Vink_2007,Curry_2015,Tracy_2016}.
For single-shot readout, qubit state distinguishability with sensitivity
140 $\mu$e/$\sqrt{\textrm{Hz}}$ has been demonstrated \citep{Barthel_2010}.
However, many of these circuits require elements to be mounted at
multiple fridge stages and the use of custom on-chip components, adding
to the complexity of their implementation. Simpler amplification circuits
that use low power transistors mounted directly on the mixing chamber
stage with the qubit device thus have significant appeal \citep{Tracy_2016,Curry_2015}.
For example, a proof of principle readout demonstration with a dual
stage HEMT achieved Te = 240 mK, gain = 2700 A/A, power = 13 $\mu$W,
noise referred to input $\leq$ 70 fA/$\sqrt{\textrm{Hz}}$, and 350
$\mu$e/$\sqrt{\textrm{Hz}}$ charge sensitivity \citep{Tracy_2016}.

Silicon-germanium (SiGe) heterojunction bipolar transistors (HBTs)
have been demonstrated to operate at liquid helium temperatures \citep{Joseph_1995,Curry_2015}
as well as millikelvin temperatures in dilution refrigerators \citep{Najafizedeh_2009,Curry_2016,Ying_2017,Davidovic_2017}.
The HBT is motivated by low 1/f noise, high R\textsubscript{out},
and possible opportunities to achieve higher gain at the same power.
Furthermore, there can be bipolar junction transistor (BJT) advantages
compared to field effect transistors (FETs) for low input impedance
amplifier circuits \citep{Horowitz_1980}. Our approach is to use
a single SiGe HBT as a cryogenic amplifier at the mixing chamber stage
of a dilution refrigerator to improve the SNR and BW of the signal
from a SET used as a charge-sensor. We have designed and characterized
two different HBT circuits: 1) the current-biased HBT circuit (CB-HBT)
(Figure \ref{fig:cb-hbt_circuit_diagram}(a)) and 2) the AC-coupled
HBT circuit (AC-HBT) (Figure \ref{fig:ac-hbt_circuit_diagram}(a)).
The CB-HBT simply has the drain of the SET connected to the base of
the HBT, while the AC-HBT has the base of the HBT connected to the
drain of the SET via a resistor-capacitor (RC) bias tee. Regardless
of the coupling between the HBT and SET, the HBT must be DC biased
in order to amplify. For either circuit, the silicon metal oxide semiconductor
(Si-MOS) device and HBT are mounted on a printed circuit board (PCB)
only centimeters apart. The proximity of the HBT amplifier to the
SET has the advantages of minimizing parasitic input capacitance and
increasing signal before noise from the fridge is added. However,
since the mixing chamber stage has a cooling power of around 100 $\mu$W
at 100 mK, the HBT circuits must operate with powers similar or less
in order to avoid heating.

In this letter, we first introduce the two amplification circuits
with discussions of gain, sensitivity, bias behavior, and noise. We
compare the basic performance and operation of the two amplifiers
and extract input-referred noise as well as signal response and heating
of the quantum dot electrons. Finally, we compare and discuss the
performance for single-shot readout, which somewhat depends on the
specific layout of the SET and quantum dot to produce larger signals
via increased mutual capacitance.

\section*{AC-HBT Description}

The AC-HBT consists of a Si-MOS device that is AC-coupled to an HBT,
which amplifies the SET response to AC source-drain voltage excitation
at frequencies higher than around 100 Hz. The SET is integrated into
a double quantum dot (QD) device (Figure \ref{fig:ac-hbt_circuit_diagram}(a):
SEM image), which is made on a Si-MOS platform (see Appendix \ref{sec:SET_Geometries}). 

To operate the AC-HBT, the DC base bias is grounded, and the emitter
is biased negatively to support a base-emitter bias $V_{BE}$ above
the cryogenic HBT threshold (about -1.04 V). The HBT current at the
collector is measured through a room temperature transimpedance amplifier
(TIA), and the signal is demodulated, filtered, and digitized. The
TIA is referenced to ground, so the collector-emitter bias equals
the base-emitter bias. We find that this configuration optimizes the
circuit SNR and also requires only two lines coming from room temperature
for the three HBT terminals. Figure \ref{fig:ac-hbt_circuit_diagram}(c)
shows the total AC circuit gain and sensitivity vs. the amount of
power dissipated by the HBT. The AC gain is measured by comparing
the current of a Coulomb blockade (CB) peak with and without the HBT.
The SET current can be measured directly by connecting the output
of $\mathrm{R_{S}}$ to a room temperature TIA (lowest ground in Figure
\ref{fig:ac-hbt_circuit_diagram}(a)). The sensitivity of the circuit
is defined as the gate-voltage derivative of collector-current (slope)
on the side of a CB peak, which is the typical bias point where readout
occurs. Sensitivities of 1-5 $\mu$A/V are achieved in the operating
region of the AC-HBT. Since the AC-HBT is a linear amplifier, the
shape of a CB peak remains unaffected by different gain/sensitivity
bias points of the AC-HBT (Figure \ref{fig:ac-hbt_circuit_diagram}(d)).
The AC bias across the SET was chosen to be 200 $\mu$V\textsubscript{RMS}
in this case to minimize the electron temperature below 200 mK.

Noise spectra are collected for different AC-HBT biases (see Appendix
\ref{sec:Noise_Models}), and noise at around 74 kHz is referred to
the HBT collector and studied. The noise displays two different behaviors
as power dissipated is increased (Figure \ref{fig:ac-hbt_circuit_diagram}(e)).
The transconductance of the transistor ($\frac{dI_{C}}{dV_{BE}}$)
increases with power, so it is important to identify where the transistor
begins to add appreciable noise. In the low-power limit, the noise
dependence is approximately flat at around 1 pA/$\sqrt{\textrm{Hz}}$,
which we attribute to the noise after the HBT dominating any AC-HBT
noise. As the AC-HBT power is increased to > 1 $\mu$W, the noise
becomes linearly dependent on power. This behavior is predicted by
our estimated shot noise for the base current (Figure \ref{fig:ac-hbt_circuit_diagram}(e)
orange curve). The estimated total noise is calculated by adding all
noise source predictions in quadrature (dark red curve) and aligns
well with the total measured noise (blue points). 

\begin{figure}
\begin{centering}
\includegraphics[width=8.5cm]{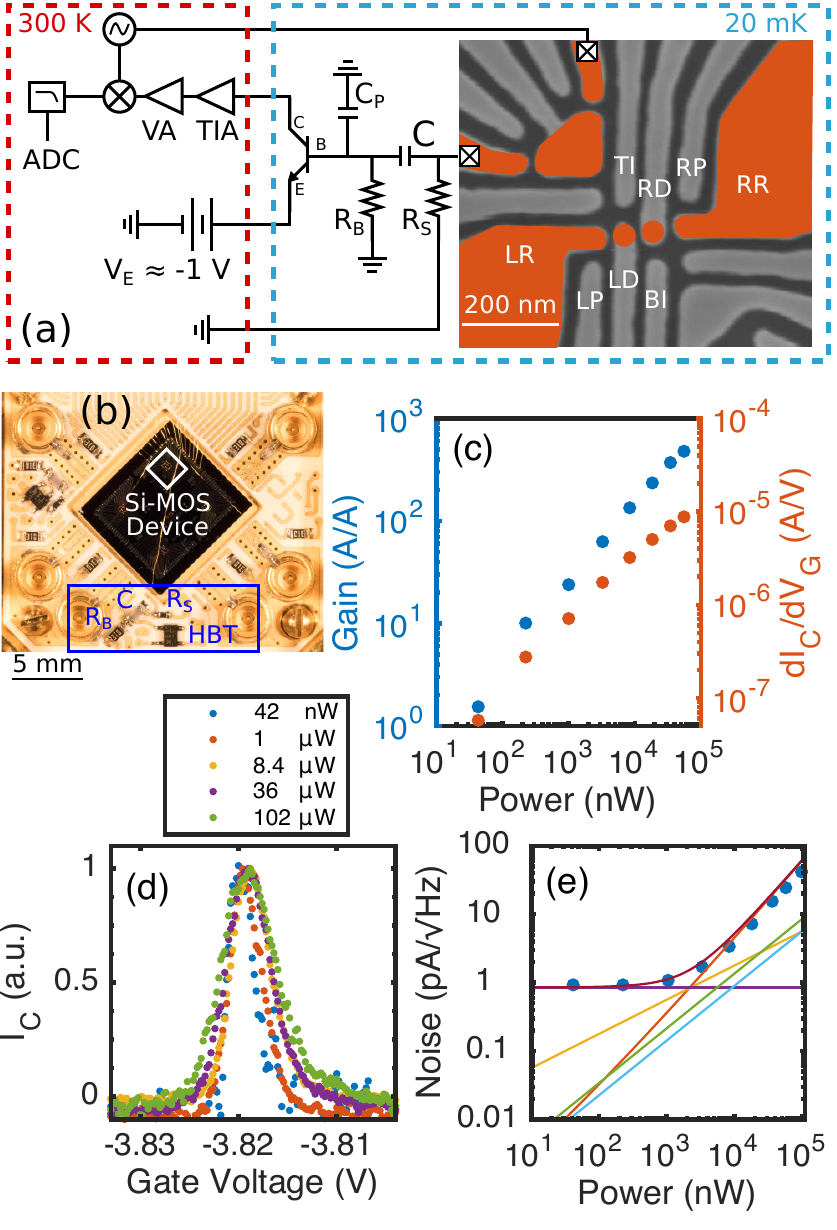}
\par\end{centering}
\caption{(a) Schematic diagram of AC-HBT and SEM image of the double quantum
dot (DQD) device. Areas of electron accumulation are indicated by
false color highlighting of enhancement gates. The charge sensor used
to measure the DQD state is in the upper left quadrant, whose source
is connected to an AC+DC signal generator, and whose drain is connected
to a cryogenic AC-coupled HBT amplification stage. The amplification
stage is mounted at the dilution refrigerator mixing chamber on the
same printed circuit board as the DQD device. Values of the passive
elements are R\protect\textsubscript{B} = 1 M\textgreek{W}, R\protect\textsubscript{S}
= 100 k\textgreek{W}, and C = 10 nF. (c) Circuit gain and sensitivity
vs. power dissipated by the AC-HBT. (d) Normalized CB peak for different
AC-HBT gain/power biases. The the blockade region of the CB peak reaches
zero current. (e) Noise referred to the collector of the AC-HBT for
different powers. The measured noise is plotted as blue points. The
noise floor of the fridge (purple), shot noise of the base (orange),
collector (yellow), SET (light blue), Johnson noise of the shunt resistor
(green), and total estimated noise (dark red) are plotted as solid
lines.}

\label{fig:ac-hbt_circuit_diagram}
\end{figure}

\section*{CB-HBT Description}

\begin{figure}
\begin{centering}
\includegraphics[width=8.5cm]{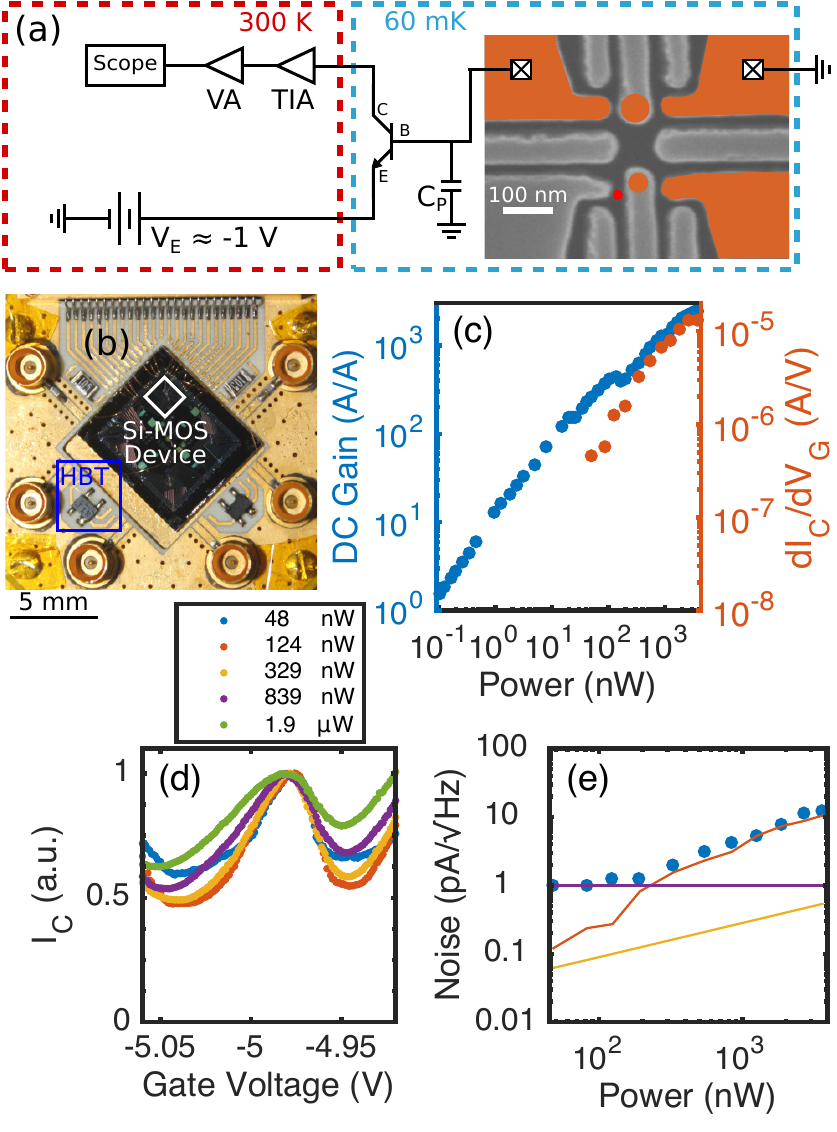}
\par\end{centering}
\caption{(a) Schematic diagram of CB-HBT readout circuit including room temperature
amplification and biasing. The SET is represented by the larger, upper
orange circle, and the QD is represented by the smaller, lower orange
circle. (b) Image of the PCB which shows the Si-MOS device and HBT
mounted close together. (c) DC current gain and sensitivity vs. power
dissipated by the CB-HBT. (d) Normalized CB peak for different CB-HBT
gain/power biases. The blockaded regions of the CB peak do not reach
zero current. (e) Noise referred to the collector of the CB-HBT for
different powers. The measured noise is plotted as blue points. For
comparison, the noise floor of the fridge (purple curve), base current
shot noise (orange curve), and collector current shot noise (yellow
curve) are plotted as well.}

\label{fig:cb-hbt_circuit_diagram}
\end{figure}

The CB-HBT circuit consists of an HBT wire bonded from its base terminal
directly to the drain of the SET. The SET is integrated into a double
QD system consisting of a lithographic QD and a secondary object that
has not been definitively identified (i.e., either a QD \citep{Jock_2018}
or donor \citep{Harvey-Collard_2017qubit}). A high-frequency coaxial
line is connected to the collector of the HBT which is used to measure
the readout current (Figure \ref{fig:cb-hbt_circuit_diagram}(a)).
This collector line is connected to a TIA which is set with gain 10\textsuperscript{5}
V/A and -3 dB bandwidth 400 kHz unless otherwise noted. The output
of the TIA is connected to a voltage amplifier used to limit the bandwidth
or further amplify the signal. Finally, the output of the voltage
amplifier is connected to an oscilloscope with an adjustable sample
rate.

Operation of the circuit requires the emitter of the HBT to be connected
to a room temperature power supply filtered to 1 MHz (to suppress
higher frequency noise) and biased between -1.03 and -1.07 V. The
bias of the emitter power supply sets the base current, collector
current, gain, and dissipated power of the HBT. In Figure \ref{fig:cb-hbt_circuit_diagram}(b),
the DC current gain and sensitivity are plotted as a function of power.
The DC current gain is defined as $\frac{I_{C}}{I_{B}}$, and the
sensitivity is defined as before. The sensitivity of the CB-HBT can
reach 5 $\mu$A/V between 100-500 nW, whereas the AC-HBT requires
> 10 $\mu$W to reach a similar sensitivity.

The CB-HBT acts as a current bias, so there is always current through
the SET (see Appendix \ref{sec:CB-HBT_Biasing_Effect}). In regions
of Coulomb blockade, the HBT base-emitter voltage will shift on the
order of the charging energy of the SET in order to maintain a relatively
constant current through the circuit. To show the current-biasing
effect, a CB peak is plotted for different CB-HBT gain values in Figure
\ref{fig:cb-hbt_circuit_diagram}(d), and the current is normalized
to the value at the top of the CB peak. Although the current in the
blockaded regions of the CB peak is much different from a voltage-biased
configuration, the slope of the sides of the CB peak appear to be
less affected by the current-biasing (sensitivities of 1--5 $\mu$A/V
are achieved for either circuit). We note that the effect of current
bias on Coulomb blockade is independent of the HBT presence (Figure
\ref{fig:cb_effect_modeling}).

As with the AC-HBT, the noise referred to the collector of the CB-HBT
is examined at around 7 kHz (Figure \ref{fig:cb-hbt_circuit_diagram}(d)).
Similar qualitatively, the lower power region is dominated by noise
after the HBT around 1 pA/$\sqrt{\textrm{Hz}}$ (purple curve). As
power is increased, the measured noise (blue points) begins to increase,
which follows the estimated behavior of the base current shot-noise
(orange curve) (see Appendix \ref{sec:Noise_Models}).

\section*{Amplifier Performance Comparison}

\begin{figure}
\begin{centering}
\includegraphics[width=8.5cm]{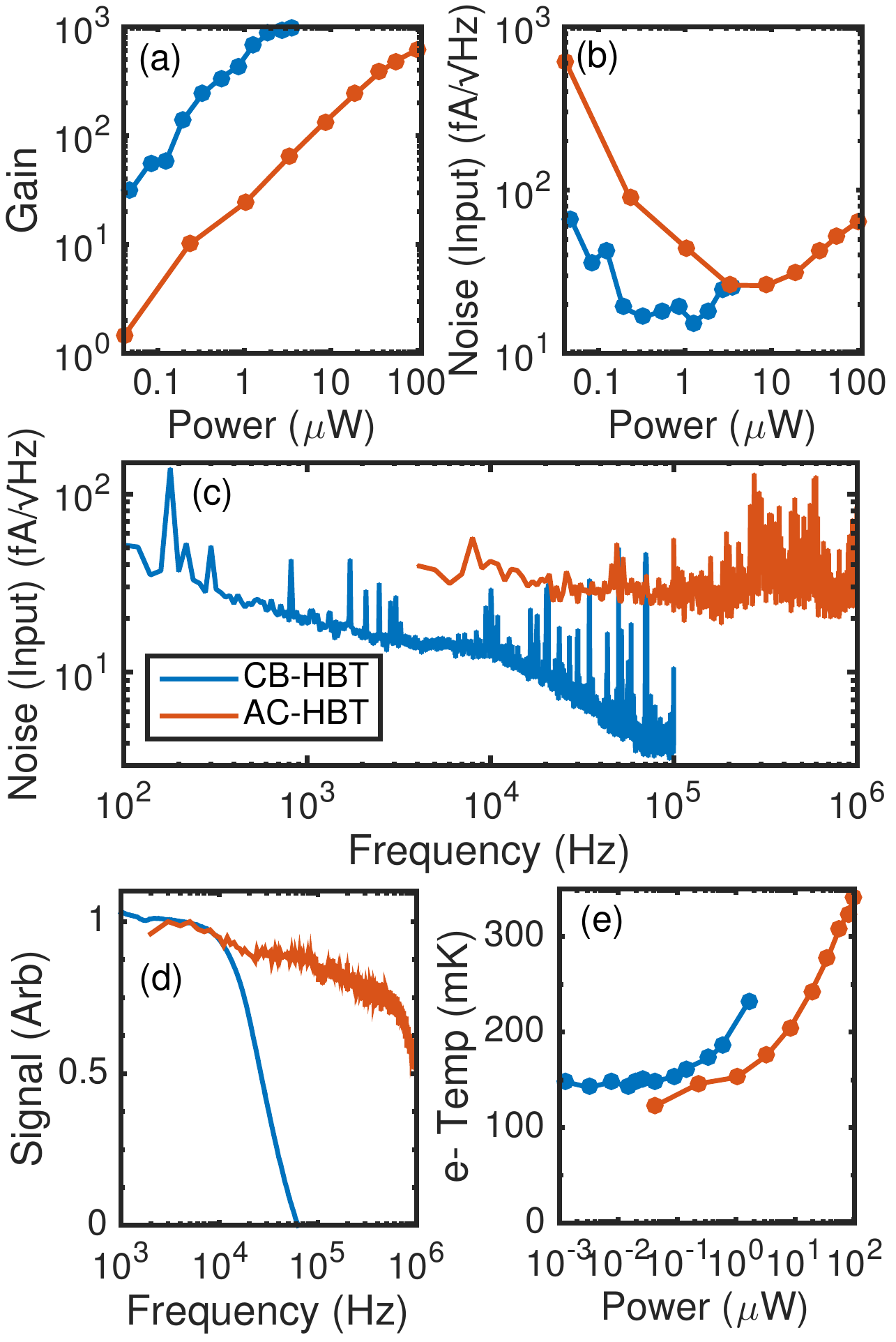}
\par\end{centering}
\caption{(a) Gain of both circuits as a function of power. Calculated gain
of the CB-HBT is shown (Appendix \ref{sec:CB-HBT_Effective_Gain}).
(b) Minimum input-referred noise as a function of power. CB-HBT has
minimum of 19 fA/$\sqrt{\textrm{Hz}}$ at 800 nW, and AC-HBT has minimum
of 26 fA/$\sqrt{\textrm{Hz}}$ at 8.4 $\mu$W. (c) Input-referred
noise spectrum of both circuits for power that minimizes noise. (d)
Signal response (in normalized arbitrary units) for both circuits
as a function of frequency. The CB-HBT has a -3 dB point at around
20 kHz, and the AC-HBT has a -3 dB point at around 650 kHz. (e) Electron
temperature vs. power for both circuits. Base electron temperatures
are between 120--150 mK.}

\label{fig:amp_comparison}
\end{figure}

We next compare the performance of both amplifiers with respect to
power dissipated. The first metric examined is gain as power is increased.
The gain of the AC-HBT is simply the measured gain of the amplifier,
however the gain of the CB-HBT circuit is not as simple to extract.
The small-signal resistance of the SET (r\textsubscript{set}) must
be known in order to calculate the CB-HBT gain (see Appendix \ref{sec:CB-HBT_Effective_Gain}).
Since the SET is directly connected to the HBT, we cannot measure
r\textsubscript{set}. Instead, we use the value of r\textsubscript{set}
(3 M\textgreek{W}) which best follows the measured noise behavior
in Figure \ref{fig:cb-hbt_circuit_diagram}(e) to estimate the gain.
We plot this estimated gain of the CB-HBT circuit and compare to the
measured gain of the AC-HBT circuit in Figure \ref{fig:amp_comparison}(a).
We observe that the CB-HBT circuit achieves higher gain at lower powers,
including operating with gain over 400 at a power around 1 $\mu$W.

We next compare the noise referred to the input of the HBT for each
circuit. Noise is referred to the input using the gain values in Figure
\ref{fig:amp_comparison}(a). We measure the noise spectrum for each
circuit at different bias points and choose the frequency which minimizes
the noise. The frequency chosen for the AC-HBT circuit was around
74 kHz, and the frequency for the CB-HBT circuit was around 7 kHz.
When the input-referred noise is plotted as a function of power (Figure
\ref{fig:amp_comparison}(b)), we observe a minimum noise operating
point for either circuit. At low powers, the noise is likely dominated
by triboelectric noise due to the fridge and input noise of the room
temperature TIA. At higher powers, the HBT amplifiers begin injecting
appreciable noise into the circuit, therefore the overall noise increases.
The CB-HBT circuit achieves a minimum noise of 19 fA/$\sqrt{\textrm{Hz}}$
at a power around 800 nW, while the AC-HBT circuit achieves a minimum
noise of 26 fA/$\sqrt{\textrm{Hz}}$ at a power around 8.4 $\mu$W.

For the powers that minimize noise for each circuit, we plot the input-referred
noise spectrum for both circuits as a function of frequency (Figure
\ref{fig:amp_comparison}(c)). The noise spectrum of the CB-HBT is
plotted out to 100 kHz, since its bandwidth is less than 100 kHz.
The 1/f-like behavior of the noise at lower frequencies is assumed
to be due to charge noise in the Si-MOS device. In the overlapping
region around 10 kHz, the noise for the CB-HBT is significantly lower
than the noise for the AC-HBT.

Figure \ref{fig:amp_comparison}(d) shows the frequency dependence
of an input signal for both amplification circuits up to 1 MHz. The
AC-HBT has a -3 dB point at around 650 kHz, and the CB-HBT has a -3
dB point at around 20 kHz, which implies significantly lower BW than
the AC-HBT. The origin of this lower BW is not well understood. Using
pessimistic numbers, the frequency pole of the SET resistance (assuming
1 M\textgreek{W}) and the parasitic capacitance between the SET and
the base junction (assuming 1 pF) should only limit the -3 dB point
to around 160 kHz. In addition, 4 K simulations of this circuit also
yielded around 160 kHz -3 dB BW \citep{England_2017}. Improvements
and understanding of the BW of the CB-HBT will be important in future
work.

Heating of electrons in the QD due to the operation of the connected
HBT is a concern, therefore we examined the dependence of electron
temperature on HBT amplifier bias (Figure \ref{fig:amp_comparison}(e)).
For the CB-HBT, we find that the minimum electron temperature observed
is around 150 mK. Heating of the QD begins where the CB-HBT is operating
with over 100 gain at 100 nW, therefore the CB-HBT circuit can amplify
well with an electron temperature around 160--200 mK. For the AC-HBT,
the minimum electron temperature was around 120 mK. When the AC-HBT
bias is increased up to 3.24 $\mu$W, the electron temperature remains
near the minimum temperature. For powers above this threshold, the
electron temperature increases approximately linearly with power.
Nonetheless, an electron temperature of 200 mK is used for the bias
condition that provides the minimum amplifier noise.

\section*{Single-Shot Results Comparison}

\begin{figure}
\begin{centering}
\includegraphics[width=8.5cm]{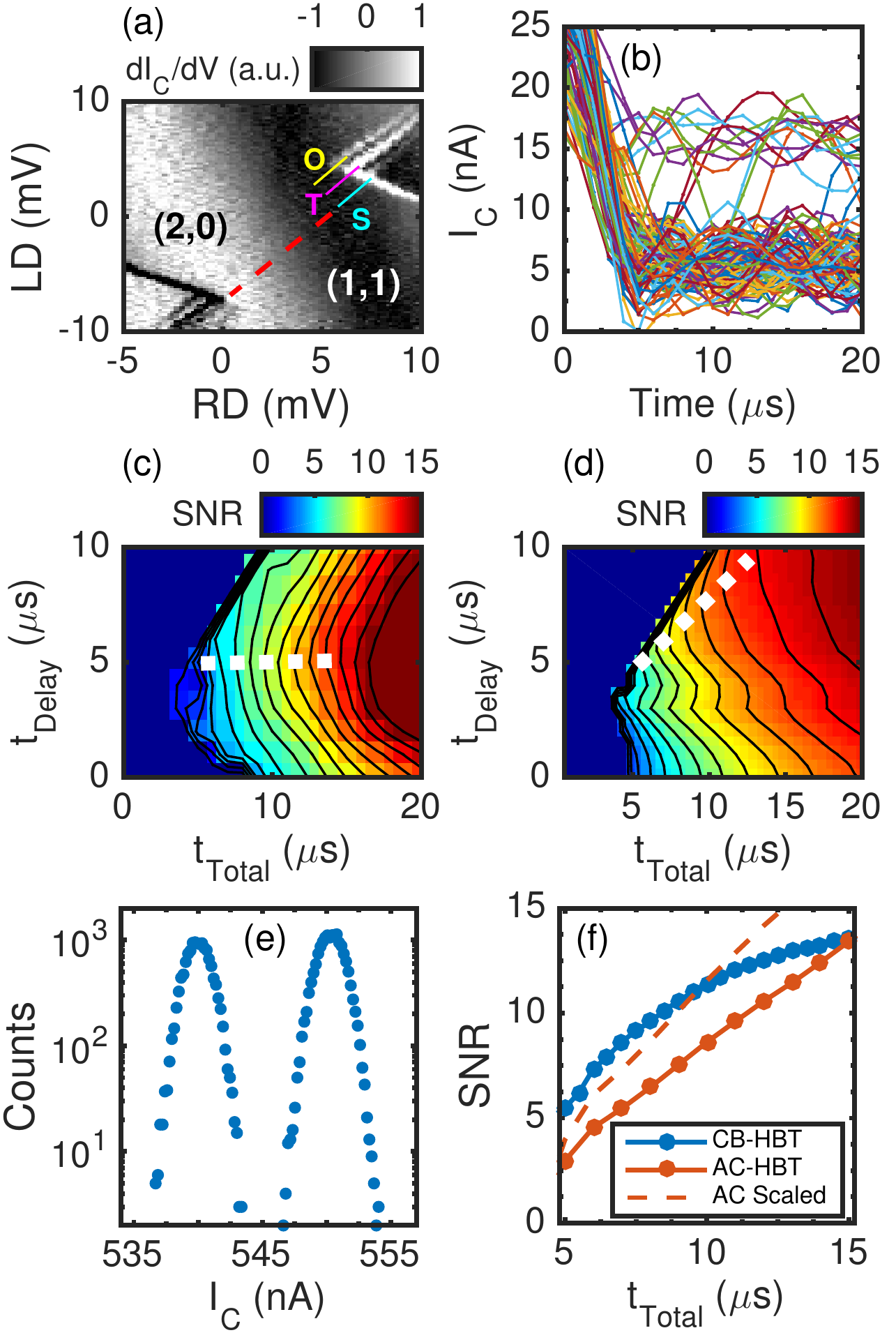}
\par\end{centering}
\caption{(a) Measurement pulse signal (derivative) rastered about the (2,0)-(1,1)
anti-crossing for the AC-HBT device. Three distinct latched lines
are present. (b) 100 single-shot traces of the readout portion of
the pulse for the AC-HBT device. Signal separation begins to occur
around 4 $\mu$s. (c) Example histogram from the CB-HBT readout. (d)
2D SNR plot for the CB-HBT readout. (e) 2D SNR plot for the AC-HBT
readout. (f) SNR vs. minimum total measurement time for both circuits,
which corresponds to the white dashed line in (c) and (d). The greater
gain of the CB-HBT compensates for the lower bandwidth relative to
the AC-HBT. The AC-HBT is also shown scaled by 34\% to compare more
directly to the CB-HBT, which had a larger SET signal.}

\label{fig:single_shot_comparison}
\end{figure}

We compare both HBT amplifiers by performing single-shot readout of
latched charge states \citep{Harvey-Collard_2018readout}. Both Si-MOS
quantum dot devices are tuned to the few electron regime and the spin
filling of the last few transition lines are verified with magnetospectroscopy.
Figure \ref{fig:single_shot_comparison}(a) shows the result of a
three-level pulse sequence in the AC-HBT device where: 1) the system
is initialized into (1,0), 2) ground and excited states are loaded
in (2,0), and finally 3) the measurement point (signal plotted) is
rastered about the (2,0)-(1,1) anti-crossing. When measuring for 30
$\mu$s, three latched lines are present, which indicates spin blockade
for an excited state triplet (T), a second excited state triplet (O),
and a lifting of the spin blockade for the ground state singlet (S).
We assign T as a valley triplet with valley splitting of 140 $\mu$eV
and the O as an orbital triplet with orbital splitting of 280 $\mu$eV.
For all single-shot measurements, we remove the state O from the available
state space by energy selective loading of the (2,0) state.

For both circuits, a mixture of (2,1) and (2,0) charge states are
read out in the reverse latching window. Figure \ref{fig:single_shot_comparison}(b)
shows 100 individual single-shot traces of the readout portion of
the pulse for the AC-HBT device. Significant feedthrough is observed
in the first few $\mu$s of the readout pulse, likely due to attenuators
connecting the conductor of the high BW lines to the ground of other
lines including the emitter bias line. State distinguishability does
not begin to occur until about 4 $\mu$s, and then the pulse relaxes
to two distinct states after about 7 $\mu$s. Extracting the SNR from
these traces is done by waiting a certain amount of time, t\textsubscript{delay},
and then averaging the signal for a certain amount of time, t\textsubscript{integration}.
Histograms of the delayed and averaged shots are compiled and fit
to a double Gaussian distribution (Figure \ref{fig:single_shot_comparison}(c)).
The signal is defined as the separation of the Gaussian peaks and
the noise is defined as the average of the widths of the Gaussian
peaks.

The extracted SNR for a given delay and total time (t\textsubscript{delay}
+ t\textsubscript{integration}) is plotted in Figures \ref{fig:single_shot_comparison}(d)\&(e).
Contours are drawn for each SNR integer on both plots, where the leftmost
part of a contour line reveals the minimum total measurement time
required to reach a given SNR. We plot the SNR vs. minimum total measurement
time in Figure \ref{fig:single_shot_comparison}(f) for both circuits.
The CB-HBT reaches greater SNR at any given time in the 15 $\mu$s
plot range. Both circuits achieve SNR > 7 in t\textsubscript{total}
< 10 $\mu$s, which corresponds to a bit error rate < 10\textsuperscript{-3}
and marks a significant improvement over the equivalent t\textsubscript{total}
= 65 $\mu$s in previous work \citep{Harvey-Collard_2018readout}.
In particular, the CB-HBT is able to reach SNR > 7 in t\textsubscript{total}
\ensuremath{\approx} 6 $\mu$s, which represents over a factor of
ten improvement from the previous work \citep{Harvey-Collard_2018readout}.
The charge sensitivity for the CB-HBT is 330 $\mu$e/$\sqrt{\textrm{Hz}}$
(\textgreek{t}\textsubscript{int} = 6 $\mu$s, SNR = 7.4), and the
charge sensitivity for the AC-HBT is 400 $\mu$e/$\sqrt{\textrm{Hz}}$
(\textgreek{t}\textsubscript{int} = 9 $\mu$s, SNR = 7.5). We note
that the SET in the CB-HBT device had around 34\% more signal due
to larger mutual capacitance (Appendix \ref{sec:SET_Geometries})
which may contribute to the larger SNRs.

The AC-HBT requires more relative overhead for implementation than
the CB-HBT. The AC-HBT includes three additional surface mounted passive
elements (Figure \ref{fig:ac-hbt_circuit_diagram}(a)), which can
be optimized to produce better SNR. Additionally, the AC-HBT has a
two-dimensional bias space via the base bias and emitter bias, whereas
the CB-HBT is only biased via the emitter bias. However, the AC-HBT
is a linear gain circuit and can be used with discrete HEMTs \citep{Tracy_2018}
and HBTs, providing more opportunity to optimize the transistor. Ideally,
the transistors would have greater transconductance (g\textsubscript{m})
and a more ideal dependence on I\textsubscript{C} than the HBTs used
in this work (see Appendix \ref{sec:HBT_Characterization}). In the
present demonstration of the AC-HBT, heating of electrons occurred
at powers which minimized noise. Introducing a second AC-HBT stage
is relatively straightforward and may allow the first stage to run
at powers which don't heat the electrons and minimize the noise further.
In addition, the second stage could be mounted further away from the
Si-MOS PCB and reduce local heating.

\section*{Conclusion}

We compare the performance of two cryogenic amplification circuits:
the CB-HBT and the AC-HBT. The power dissipated by the CB-HBT ranges
from 0.1 to 1 $\mu$W, whereas the power of the AC-HBT ranges from
1 to 20 $\mu$W. Referred to the input, the noise spectral density
is low for both circuits in the 15 to 30 fA/$\sqrt{\textrm{Hz}}$
range. The charge sensitivity for the CB-HBT and AC-HBT is 330 $\mu$e/$\sqrt{\textrm{Hz}}$
and 400 $\mu$e/$\sqrt{\textrm{Hz}}$, respectively. For single-shot
readout performed, both circuits achieve SNR > 7 and bit error rate
< 10\textsuperscript{-3} in times less than 10 $\mu$s.
\begin{acknowledgments}
This work was performed, in part, at the Center for Integrated Nanotechnologies,
an Office of Science User Facility operated for the U.S. Department
of Energy (DOE) Office of Science. Sandia National Laboratories is
a multi-mission laboratory managed and operated by National Technology
and Engineering Solutions of Sandia, LLC, a wholly owned subsidiary
of Honeywell International, Inc., for the DOE\textquoteright s National
Nuclear Security Administration under contract DE-NA0003525.

This paper describes objective technical results and analysis. Any
subjective views or opinions that might be expressed in the paper
do not necessarily represent the views of the U.S. Department of Energy
or the United States Government.
\end{acknowledgments}


\begin{thebibliography}{52}%
\makeatletter
\providecommand \@ifxundefined [1]{%
 \@ifx{#1\undefined}
}%
\providecommand \@ifnum [1]{%
 \ifnum #1\expandafter \@firstoftwo
 \else \expandafter \@secondoftwo
 \fi
}%
\providecommand \@ifx [1]{%
 \ifx #1\expandafter \@firstoftwo
 \else \expandafter \@secondoftwo
 \fi
}%
\providecommand \natexlab [1]{#1}%
\providecommand \enquote  [1]{``#1''}%
\providecommand \bibnamefont  [1]{#1}%
\providecommand \bibfnamefont [1]{#1}%
\providecommand \citenamefont [1]{#1}%
\providecommand \href@noop [0]{\@secondoftwo}%
\providecommand \href [0]{\begingroup \@sanitize@url \@href}%
\providecommand \@href[1]{\@@startlink{#1}\@@href}%
\providecommand \@@href[1]{\endgroup#1\@@endlink}%
\providecommand \@sanitize@url [0]{\catcode `\\12\catcode `\$12\catcode
  `\&12\catcode `\#12\catcode `\^12\catcode `\_12\catcode `\%12\relax}%
\providecommand \@@startlink[1]{}%
\providecommand \@@endlink[0]{}%
\providecommand \url  [0]{\begingroup\@sanitize@url \@url }%
\providecommand \@url [1]{\endgroup\@href {#1}{\urlprefix }}%
\providecommand \urlprefix  [0]{URL }%
\providecommand \Eprint [0]{\href }%
\providecommand \doibase [0]{http://dx.doi.org/}%
\providecommand \selectlanguage [0]{\@gobble}%
\providecommand \bibinfo  [0]{\@secondoftwo}%
\providecommand \bibfield  [0]{\@secondoftwo}%
\providecommand \translation [1]{[#1]}%
\providecommand \BibitemOpen [0]{}%
\providecommand \bibitemStop [0]{}%
\providecommand \bibitemNoStop [0]{.\EOS\space}%
\providecommand \EOS [0]{\spacefactor3000\relax}%
\providecommand \BibitemShut  [1]{\csname bibitem#1\endcsname}%
\let\auto@bib@innerbib\@empty
\bibitem [{\citenamefont {Kane}(1998)}]{Kane_1998}%
  \BibitemOpen
  \bibfield  {author} {\bibinfo {author} {\bibfnamefont {B.~E.}\ \bibnamefont
  {Kane}},\ }\href {http://dx.doi.org/10.1038/30156} {\bibfield  {journal}
  {\bibinfo  {journal} {Nature}\ }\textbf {\bibinfo {volume} {393}},\ \bibinfo
  {pages} {133 EP } (\bibinfo {year} {1998})}\BibitemShut {NoStop}%
\bibitem [{\citenamefont {Elzerman}\ \emph {et~al.}(2004)\citenamefont
  {Elzerman}, \citenamefont {Hanson}, \citenamefont {Willems~van Beveren},
  \citenamefont {Witkamp}, \citenamefont {Vandersypen},\ and\ \citenamefont
  {Kouwenhoven}}]{Elzerman_2004}%
  \BibitemOpen
  \bibfield  {author} {\bibinfo {author} {\bibfnamefont {J.~M.}\ \bibnamefont
  {Elzerman}}, \bibinfo {author} {\bibfnamefont {R.}~\bibnamefont {Hanson}},
  \bibinfo {author} {\bibfnamefont {L.~H.}\ \bibnamefont {Willems~van
  Beveren}}, \bibinfo {author} {\bibfnamefont {B.}~\bibnamefont {Witkamp}},
  \bibinfo {author} {\bibfnamefont {L.~M.~K.}\ \bibnamefont {Vandersypen}}, \
  and\ \bibinfo {author} {\bibfnamefont {L.~P.}\ \bibnamefont {Kouwenhoven}},\
  }\href {http://dx.doi.org/10.1038/nature02693} {\bibfield  {journal}
  {\bibinfo  {journal} {Nature}\ }\textbf {\bibinfo {volume} {430}},\ \bibinfo
  {pages} {431 EP } (\bibinfo {year} {2004})}\BibitemShut {NoStop}%
\bibitem [{\citenamefont {Petta}\ \emph {et~al.}(2005)\citenamefont {Petta},
  \citenamefont {Johnson}, \citenamefont {Taylor}, \citenamefont {Laird},
  \citenamefont {Yacoby}, \citenamefont {Lukin}, \citenamefont {Marcus},
  \citenamefont {Hanson},\ and\ \citenamefont {Gossard}}]{Petta_2005}%
  \BibitemOpen
  \bibfield  {author} {\bibinfo {author} {\bibfnamefont {J.~R.}\ \bibnamefont
  {Petta}}, \bibinfo {author} {\bibfnamefont {A.~C.}\ \bibnamefont {Johnson}},
  \bibinfo {author} {\bibfnamefont {J.~M.}\ \bibnamefont {Taylor}}, \bibinfo
  {author} {\bibfnamefont {E.~A.}\ \bibnamefont {Laird}}, \bibinfo {author}
  {\bibfnamefont {A.}~\bibnamefont {Yacoby}}, \bibinfo {author} {\bibfnamefont
  {M.~D.}\ \bibnamefont {Lukin}}, \bibinfo {author} {\bibfnamefont {C.~M.}\
  \bibnamefont {Marcus}}, \bibinfo {author} {\bibfnamefont {M.~P.}\
  \bibnamefont {Hanson}}, \ and\ \bibinfo {author} {\bibfnamefont {A.~C.}\
  \bibnamefont {Gossard}},\ }\href {\doibase 10.1126/science.1116955}
  {\bibfield  {journal} {\bibinfo  {journal} {Science}\ }\textbf {\bibinfo
  {volume} {309}},\ \bibinfo {pages} {2180} (\bibinfo {year}
  {2005})}\BibitemShut {NoStop}%
\bibitem [{\citenamefont {Morello}\ \emph {et~al.}(2010)\citenamefont
  {Morello}, \citenamefont {Pla}, \citenamefont {Zwanenburg}, \citenamefont
  {Chan}, \citenamefont {Tan}, \citenamefont {Huebl}, \citenamefont
  {M{\"o}tt{\"o}nen}, \citenamefont {Nugroho}, \citenamefont {Yang},
  \citenamefont {van Donkelaar}, \citenamefont {Alves}, \citenamefont
  {Jamieson}, \citenamefont {Escott}, \citenamefont {Hollenberg}, \citenamefont
  {Clark},\ and\ \citenamefont {Dzurak}}]{Morello_2010}%
  \BibitemOpen
  \bibfield  {author} {\bibinfo {author} {\bibfnamefont {A.}~\bibnamefont
  {Morello}}, \bibinfo {author} {\bibfnamefont {J.~J.}\ \bibnamefont {Pla}},
  \bibinfo {author} {\bibfnamefont {F.~A.}\ \bibnamefont {Zwanenburg}},
  \bibinfo {author} {\bibfnamefont {K.~W.}\ \bibnamefont {Chan}}, \bibinfo
  {author} {\bibfnamefont {K.~Y.}\ \bibnamefont {Tan}}, \bibinfo {author}
  {\bibfnamefont {H.}~\bibnamefont {Huebl}}, \bibinfo {author} {\bibfnamefont
  {M.}~\bibnamefont {M{\"o}tt{\"o}nen}}, \bibinfo {author} {\bibfnamefont
  {C.~D.}\ \bibnamefont {Nugroho}}, \bibinfo {author} {\bibfnamefont
  {C.}~\bibnamefont {Yang}}, \bibinfo {author} {\bibfnamefont {J.~A.}\
  \bibnamefont {van Donkelaar}}, \bibinfo {author} {\bibfnamefont {A.~D.~C.}\
  \bibnamefont {Alves}}, \bibinfo {author} {\bibfnamefont {D.~N.}\ \bibnamefont
  {Jamieson}}, \bibinfo {author} {\bibfnamefont {C.~C.}\ \bibnamefont
  {Escott}}, \bibinfo {author} {\bibfnamefont {L.~C.~L.}\ \bibnamefont
  {Hollenberg}}, \bibinfo {author} {\bibfnamefont {R.~G.}\ \bibnamefont
  {Clark}}, \ and\ \bibinfo {author} {\bibfnamefont {A.~S.}\ \bibnamefont
  {Dzurak}},\ }\href {http://dx.doi.org/10.1038/nature09392} {\bibfield
  {journal} {\bibinfo  {journal} {Nature}\ }\textbf {\bibinfo {volume} {467}},\
  \bibinfo {pages} {687 EP } (\bibinfo {year} {2010})}\BibitemShut {NoStop}%
\bibitem [{\citenamefont {Pla}\ \emph {et~al.}(2013)\citenamefont {Pla},
  \citenamefont {Tan}, \citenamefont {Dehollain}, \citenamefont {Lim},
  \citenamefont {Morton}, \citenamefont {Zwanenburg}, \citenamefont {Jamieson},
  \citenamefont {Dzurak},\ and\ \citenamefont {Morello}}]{Pla_2013}%
  \BibitemOpen
  \bibfield  {author} {\bibinfo {author} {\bibfnamefont {J.~J.}\ \bibnamefont
  {Pla}}, \bibinfo {author} {\bibfnamefont {K.~Y.}\ \bibnamefont {Tan}},
  \bibinfo {author} {\bibfnamefont {J.~P.}\ \bibnamefont {Dehollain}}, \bibinfo
  {author} {\bibfnamefont {W.~H.}\ \bibnamefont {Lim}}, \bibinfo {author}
  {\bibfnamefont {J.~J.~L.}\ \bibnamefont {Morton}}, \bibinfo {author}
  {\bibfnamefont {F.~A.}\ \bibnamefont {Zwanenburg}}, \bibinfo {author}
  {\bibfnamefont {D.~N.}\ \bibnamefont {Jamieson}}, \bibinfo {author}
  {\bibfnamefont {A.~S.}\ \bibnamefont {Dzurak}}, \ and\ \bibinfo {author}
  {\bibfnamefont {A.}~\bibnamefont {Morello}},\ }\href
  {http://dx.doi.org/10.1038/nature12011} {\bibfield  {journal} {\bibinfo
  {journal} {Nature}\ }\textbf {\bibinfo {volume} {496}},\ \bibinfo {pages}
  {334 EP } (\bibinfo {year} {2013})}\BibitemShut {NoStop}%
\bibitem [{\citenamefont {Eng}\ \emph {et~al.}(2015)\citenamefont {Eng},
  \citenamefont {Ladd}, \citenamefont {Smith}, \citenamefont {Borselli},
  \citenamefont {Kiselev}, \citenamefont {Fong}, \citenamefont {Holabird},
  \citenamefont {Hazard}, \citenamefont {Huang}, \citenamefont {Deelman},\ and\
  \citenamefont {et~al.}}]{Eng_2015}%
  \BibitemOpen
  \bibfield  {author} {\bibinfo {author} {\bibfnamefont {K.}~\bibnamefont
  {Eng}}, \bibinfo {author} {\bibfnamefont {T.~D.}\ \bibnamefont {Ladd}},
  \bibinfo {author} {\bibfnamefont {A.}~\bibnamefont {Smith}}, \bibinfo
  {author} {\bibfnamefont {M.~G.}\ \bibnamefont {Borselli}}, \bibinfo {author}
  {\bibfnamefont {A.~A.}\ \bibnamefont {Kiselev}}, \bibinfo {author}
  {\bibfnamefont {B.~H.}\ \bibnamefont {Fong}}, \bibinfo {author}
  {\bibfnamefont {K.~S.}\ \bibnamefont {Holabird}}, \bibinfo {author}
  {\bibfnamefont {T.~M.}\ \bibnamefont {Hazard}}, \bibinfo {author}
  {\bibfnamefont {B.}~\bibnamefont {Huang}}, \bibinfo {author} {\bibfnamefont
  {P.~W.}\ \bibnamefont {Deelman}}, \ and\ \bibinfo {author} {\bibnamefont
  {et~al.}},\ }\href {\doibase 10.1126/sciadv.1500214} {\bibfield  {journal}
  {\bibinfo  {journal} {Science Advances}\ }\textbf {\bibinfo {volume} {1}},\
  \bibinfo {pages} {e1500214} (\bibinfo {year} {2015})}\BibitemShut {NoStop}%
\bibitem [{\citenamefont {Zajac}\ \emph {et~al.}(2017)\citenamefont {Zajac},
  \citenamefont {Sigillito}, \citenamefont {Russ}, \citenamefont {Borjans},
  \citenamefont {Taylor}, \citenamefont {Burkard},\ and\ \citenamefont
  {Petta}}]{Zajac_2017}%
  \BibitemOpen
  \bibfield  {author} {\bibinfo {author} {\bibfnamefont {D.~M.}\ \bibnamefont
  {Zajac}}, \bibinfo {author} {\bibfnamefont {A.~J.}\ \bibnamefont
  {Sigillito}}, \bibinfo {author} {\bibfnamefont {M.}~\bibnamefont {Russ}},
  \bibinfo {author} {\bibfnamefont {F.}~\bibnamefont {Borjans}}, \bibinfo
  {author} {\bibfnamefont {J.~M.}\ \bibnamefont {Taylor}}, \bibinfo {author}
  {\bibfnamefont {G.}~\bibnamefont {Burkard}}, \ and\ \bibinfo {author}
  {\bibfnamefont {J.~R.}\ \bibnamefont {Petta}},\ }\href {\doibase
  10.1126/science.aao5965} {\bibfield  {journal} {\bibinfo  {journal}
  {Science}\ ,\ \bibinfo {pages} {eaao5965}} (\bibinfo {year}
  {2017})}\BibitemShut {NoStop}%
\bibitem [{\citenamefont {Rochette}\ \emph {et~al.}(2017)\citenamefont
  {Rochette}, \citenamefont {Rudolph}, \citenamefont {Roy}, \citenamefont
  {Curry}, \citenamefont {Eyck}, \citenamefont {Manginell}, \citenamefont
  {Wendt}, \citenamefont {Pluym}, \citenamefont {Carr}, \citenamefont {Ward}
  \emph {et~al.}}]{Rochette_2017}%
  \BibitemOpen
  \bibfield  {author} {\bibinfo {author} {\bibfnamefont {S.}~\bibnamefont
  {Rochette}}, \bibinfo {author} {\bibfnamefont {M.}~\bibnamefont {Rudolph}},
  \bibinfo {author} {\bibfnamefont {A.-M.}\ \bibnamefont {Roy}}, \bibinfo
  {author} {\bibfnamefont {M.}~\bibnamefont {Curry}}, \bibinfo {author}
  {\bibfnamefont {G.~T.}\ \bibnamefont {Eyck}}, \bibinfo {author}
  {\bibfnamefont {R.}~\bibnamefont {Manginell}}, \bibinfo {author}
  {\bibfnamefont {J.}~\bibnamefont {Wendt}}, \bibinfo {author} {\bibfnamefont
  {T.}~\bibnamefont {Pluym}}, \bibinfo {author} {\bibfnamefont
  {S.}~\bibnamefont {Carr}}, \bibinfo {author} {\bibfnamefont {D.}~\bibnamefont
  {Ward}},  \emph {et~al.},\ }\href {https://arxiv.org/abs/1707.03895}
  {\bibfield  {journal} {\bibinfo  {journal} {arXiv preprint arXiv:1707.03895}\
  } (\bibinfo {year} {2017})}\BibitemShut {NoStop}%
\bibitem [{\citenamefont {Muhonen}\ \emph {et~al.}(2014)\citenamefont
  {Muhonen}, \citenamefont {Dehollain}, \citenamefont {Laucht}, \citenamefont
  {Hudson}, \citenamefont {Kalra}, \citenamefont {Sekiguchi}, \citenamefont
  {Itoh}, \citenamefont {Jamieson}, \citenamefont {McCallum}, \citenamefont
  {Dzurak},\ and\ \citenamefont {Morello}}]{Muhonen_2014}%
  \BibitemOpen
  \bibfield  {author} {\bibinfo {author} {\bibfnamefont {J.~T.}\ \bibnamefont
  {Muhonen}}, \bibinfo {author} {\bibfnamefont {J.~P.}\ \bibnamefont
  {Dehollain}}, \bibinfo {author} {\bibfnamefont {A.}~\bibnamefont {Laucht}},
  \bibinfo {author} {\bibfnamefont {F.~E.}\ \bibnamefont {Hudson}}, \bibinfo
  {author} {\bibfnamefont {R.}~\bibnamefont {Kalra}}, \bibinfo {author}
  {\bibfnamefont {T.}~\bibnamefont {Sekiguchi}}, \bibinfo {author}
  {\bibfnamefont {K.~M.}\ \bibnamefont {Itoh}}, \bibinfo {author}
  {\bibfnamefont {D.~N.}\ \bibnamefont {Jamieson}}, \bibinfo {author}
  {\bibfnamefont {J.~C.}\ \bibnamefont {McCallum}}, \bibinfo {author}
  {\bibfnamefont {A.~S.}\ \bibnamefont {Dzurak}}, \ and\ \bibinfo {author}
  {\bibfnamefont {A.}~\bibnamefont {Morello}},\ }\href
  {http://dx.doi.org/10.1038/nnano.2014.211} {\bibfield  {journal} {\bibinfo
  {journal} {Nature Nanotechnology}\ }\textbf {\bibinfo {volume} {9}},\
  \bibinfo {pages} {986 EP } (\bibinfo {year} {2014})}\BibitemShut {NoStop}%
\bibitem [{\citenamefont {Harvey-Collard}\ \emph {et~al.}(2018)\citenamefont
  {Harvey-Collard}, \citenamefont {D'Anjou}, \citenamefont {Rudolph},
  \citenamefont {Jacobson}, \citenamefont {Dominguez}, \citenamefont
  {Ten~Eyck}, \citenamefont {Wendt}, \citenamefont {Pluym}, \citenamefont
  {Lilly}, \citenamefont {Coish}, \citenamefont {Pioro-Ladri\`ere},\ and\
  \citenamefont {Carroll}}]{Harvey-Collard_2018readout}%
  \BibitemOpen
  \bibfield  {author} {\bibinfo {author} {\bibfnamefont {P.}~\bibnamefont
  {Harvey-Collard}}, \bibinfo {author} {\bibfnamefont {B.}~\bibnamefont
  {D'Anjou}}, \bibinfo {author} {\bibfnamefont {M.}~\bibnamefont {Rudolph}},
  \bibinfo {author} {\bibfnamefont {N.~T.}\ \bibnamefont {Jacobson}}, \bibinfo
  {author} {\bibfnamefont {J.}~\bibnamefont {Dominguez}}, \bibinfo {author}
  {\bibfnamefont {G.~A.}\ \bibnamefont {Ten~Eyck}}, \bibinfo {author}
  {\bibfnamefont {J.~R.}\ \bibnamefont {Wendt}}, \bibinfo {author}
  {\bibfnamefont {T.}~\bibnamefont {Pluym}}, \bibinfo {author} {\bibfnamefont
  {M.~P.}\ \bibnamefont {Lilly}}, \bibinfo {author} {\bibfnamefont {W.~A.}\
  \bibnamefont {Coish}}, \bibinfo {author} {\bibfnamefont {M.}~\bibnamefont
  {Pioro-Ladri\`ere}}, \ and\ \bibinfo {author} {\bibfnamefont {M.~S.}\
  \bibnamefont {Carroll}},\ }\href {\doibase 10.1103/PhysRevX.8.021046}
  {\bibfield  {journal} {\bibinfo  {journal} {Phys. Rev. X}\ }\textbf {\bibinfo
  {volume} {8}},\ \bibinfo {pages} {021046} (\bibinfo {year}
  {2018})}\BibitemShut {NoStop}%
\bibitem [{\citenamefont {Nakajima}\ \emph {et~al.}(2017)\citenamefont
  {Nakajima}, \citenamefont {Delbecq}, \citenamefont {Otsuka}, \citenamefont
  {Stano}, \citenamefont {Amaha}, \citenamefont {Yoneda}, \citenamefont
  {Noiri}, \citenamefont {Kawasaki}, \citenamefont {Takeda}, \citenamefont
  {Allison}, \citenamefont {Ludwig}, \citenamefont {Wieck}, \citenamefont
  {Loss},\ and\ \citenamefont {Tarucha}}]{Nakajima_2017}%
  \BibitemOpen
  \bibfield  {author} {\bibinfo {author} {\bibfnamefont {T.}~\bibnamefont
  {Nakajima}}, \bibinfo {author} {\bibfnamefont {M.~R.}\ \bibnamefont
  {Delbecq}}, \bibinfo {author} {\bibfnamefont {T.}~\bibnamefont {Otsuka}},
  \bibinfo {author} {\bibfnamefont {P.}~\bibnamefont {Stano}}, \bibinfo
  {author} {\bibfnamefont {S.}~\bibnamefont {Amaha}}, \bibinfo {author}
  {\bibfnamefont {J.}~\bibnamefont {Yoneda}}, \bibinfo {author} {\bibfnamefont
  {A.}~\bibnamefont {Noiri}}, \bibinfo {author} {\bibfnamefont
  {K.}~\bibnamefont {Kawasaki}}, \bibinfo {author} {\bibfnamefont
  {K.}~\bibnamefont {Takeda}}, \bibinfo {author} {\bibfnamefont
  {G.}~\bibnamefont {Allison}}, \bibinfo {author} {\bibfnamefont
  {A.}~\bibnamefont {Ludwig}}, \bibinfo {author} {\bibfnamefont {A.~D.}\
  \bibnamefont {Wieck}}, \bibinfo {author} {\bibfnamefont {D.}~\bibnamefont
  {Loss}}, \ and\ \bibinfo {author} {\bibfnamefont {S.}~\bibnamefont
  {Tarucha}},\ }\href {\doibase 10.1103/PhysRevLett.119.017701} {\bibfield
  {journal} {\bibinfo  {journal} {Phys. Rev. Lett.}\ }\textbf {\bibinfo
  {volume} {119}},\ \bibinfo {pages} {017701} (\bibinfo {year}
  {2017})}\BibitemShut {NoStop}%
\bibitem [{\citenamefont {Shulman}\ \emph {et~al.}(2014)\citenamefont
  {Shulman}, \citenamefont {Harvey}, \citenamefont {Nichol}, \citenamefont
  {Bartlett}, \citenamefont {Doherty}, \citenamefont {Umansky},\ and\
  \citenamefont {Yacoby}}]{Shulman_2014}%
  \BibitemOpen
  \bibfield  {author} {\bibinfo {author} {\bibfnamefont {M.~D.}\ \bibnamefont
  {Shulman}}, \bibinfo {author} {\bibfnamefont {S.~P.}\ \bibnamefont {Harvey}},
  \bibinfo {author} {\bibfnamefont {J.~M.}\ \bibnamefont {Nichol}}, \bibinfo
  {author} {\bibfnamefont {S.~D.}\ \bibnamefont {Bartlett}}, \bibinfo {author}
  {\bibfnamefont {A.~C.}\ \bibnamefont {Doherty}}, \bibinfo {author}
  {\bibfnamefont {V.}~\bibnamefont {Umansky}}, \ and\ \bibinfo {author}
  {\bibfnamefont {A.}~\bibnamefont {Yacoby}},\ }\href
  {http://dx.doi.org/10.1038/ncomms6156} {\bibfield  {journal} {\bibinfo
  {journal} {Nature Communications}\ }\textbf {\bibinfo {volume} {5}},\
  \bibinfo {pages} {5156 EP } (\bibinfo {year} {2014})}\BibitemShut {NoStop}%
\bibitem [{\citenamefont {Watson}\ \emph {et~al.}(2015)\citenamefont {Watson},
  \citenamefont {Weber}, \citenamefont {House}, \citenamefont {B\"uch},\ and\
  \citenamefont {Simmons}}]{Watson_2015}%
  \BibitemOpen
  \bibfield  {author} {\bibinfo {author} {\bibfnamefont {T.~F.}\ \bibnamefont
  {Watson}}, \bibinfo {author} {\bibfnamefont {B.}~\bibnamefont {Weber}},
  \bibinfo {author} {\bibfnamefont {M.~G.}\ \bibnamefont {House}}, \bibinfo
  {author} {\bibfnamefont {H.}~\bibnamefont {B\"uch}}, \ and\ \bibinfo {author}
  {\bibfnamefont {M.~Y.}\ \bibnamefont {Simmons}},\ }\href {\doibase
  10.1103/PhysRevLett.115.166806} {\bibfield  {journal} {\bibinfo  {journal}
  {Phys. Rev. Lett.}\ }\textbf {\bibinfo {volume} {115}},\ \bibinfo {pages}
  {166806} (\bibinfo {year} {2015})}\BibitemShut {NoStop}%
\bibitem [{\citenamefont {Takeda}\ \emph {et~al.}(2016)\citenamefont {Takeda},
  \citenamefont {Kamioka}, \citenamefont {Otsuka}, \citenamefont {Yoneda},
  \citenamefont {Nakajima}, \citenamefont {Delbecq}, \citenamefont {Amaha},
  \citenamefont {Allison}, \citenamefont {Kodera}, \citenamefont {Oda},\ and\
  \citenamefont {Tarucha}}]{Takeda_2016}%
  \BibitemOpen
  \bibfield  {author} {\bibinfo {author} {\bibfnamefont {K.}~\bibnamefont
  {Takeda}}, \bibinfo {author} {\bibfnamefont {J.}~\bibnamefont {Kamioka}},
  \bibinfo {author} {\bibfnamefont {T.}~\bibnamefont {Otsuka}}, \bibinfo
  {author} {\bibfnamefont {J.}~\bibnamefont {Yoneda}}, \bibinfo {author}
  {\bibfnamefont {T.}~\bibnamefont {Nakajima}}, \bibinfo {author}
  {\bibfnamefont {M.~R.}\ \bibnamefont {Delbecq}}, \bibinfo {author}
  {\bibfnamefont {S.}~\bibnamefont {Amaha}}, \bibinfo {author} {\bibfnamefont
  {G.}~\bibnamefont {Allison}}, \bibinfo {author} {\bibfnamefont
  {T.}~\bibnamefont {Kodera}}, \bibinfo {author} {\bibfnamefont
  {S.}~\bibnamefont {Oda}}, \ and\ \bibinfo {author} {\bibfnamefont
  {S.}~\bibnamefont {Tarucha}},\ }\href {\doibase 10.1126/sciadv.1600694}
  {\bibfield  {journal} {\bibinfo  {journal} {Science Advances}\ }\textbf
  {\bibinfo {volume} {2}} (\bibinfo {year} {2016}),\
  10.1126/sciadv.1600694}\BibitemShut {NoStop}%
\bibitem [{\citenamefont {Kawakami}\ \emph {et~al.}(2016)\citenamefont
  {Kawakami}, \citenamefont {Jullien}, \citenamefont {Scarlino}, \citenamefont
  {Ward}, \citenamefont {Savage}, \citenamefont {Lagally}, \citenamefont
  {Dobrovitski}, \citenamefont {Friesen}, \citenamefont {Coppersmith},
  \citenamefont {Eriksson},\ and\ \citenamefont {Vandersypen}}]{Kawakami_2016}%
  \BibitemOpen
  \bibfield  {author} {\bibinfo {author} {\bibfnamefont {E.}~\bibnamefont
  {Kawakami}}, \bibinfo {author} {\bibfnamefont {T.}~\bibnamefont {Jullien}},
  \bibinfo {author} {\bibfnamefont {P.}~\bibnamefont {Scarlino}}, \bibinfo
  {author} {\bibfnamefont {D.~R.}\ \bibnamefont {Ward}}, \bibinfo {author}
  {\bibfnamefont {D.~E.}\ \bibnamefont {Savage}}, \bibinfo {author}
  {\bibfnamefont {M.~G.}\ \bibnamefont {Lagally}}, \bibinfo {author}
  {\bibfnamefont {V.~V.}\ \bibnamefont {Dobrovitski}}, \bibinfo {author}
  {\bibfnamefont {M.}~\bibnamefont {Friesen}}, \bibinfo {author} {\bibfnamefont
  {S.~N.}\ \bibnamefont {Coppersmith}}, \bibinfo {author} {\bibfnamefont
  {M.~A.}\ \bibnamefont {Eriksson}}, \ and\ \bibinfo {author} {\bibfnamefont
  {L.~M.~K.}\ \bibnamefont {Vandersypen}},\ }\href {\doibase
  10.1073/pnas.1603251113} {\bibfield  {journal} {\bibinfo  {journal}
  {Proceedings of the National Academy of Sciences}\ }\textbf {\bibinfo
  {volume} {113}},\ \bibinfo {pages} {11738} (\bibinfo {year}
  {2016})}\BibitemShut {NoStop}%
\bibitem [{\citenamefont {Nichol}\ \emph {et~al.}(2017)\citenamefont {Nichol},
  \citenamefont {Orona}, \citenamefont {Harvey}, \citenamefont {Fallahi},
  \citenamefont {Gardner}, \citenamefont {Manfra},\ and\ \citenamefont
  {Yacoby}}]{Nichol_2017}%
  \BibitemOpen
  \bibfield  {author} {\bibinfo {author} {\bibfnamefont {J.~M.}\ \bibnamefont
  {Nichol}}, \bibinfo {author} {\bibfnamefont {L.~A.}\ \bibnamefont {Orona}},
  \bibinfo {author} {\bibfnamefont {S.~P.}\ \bibnamefont {Harvey}}, \bibinfo
  {author} {\bibfnamefont {S.}~\bibnamefont {Fallahi}}, \bibinfo {author}
  {\bibfnamefont {G.~C.}\ \bibnamefont {Gardner}}, \bibinfo {author}
  {\bibfnamefont {M.~J.}\ \bibnamefont {Manfra}}, \ and\ \bibinfo {author}
  {\bibfnamefont {A.}~\bibnamefont {Yacoby}},\ }\href {\doibase
  10.1038/s41534-016-0003-1} {\bibfield  {journal} {\bibinfo  {journal} {npj
  Quantum Information}\ }\textbf {\bibinfo {volume} {3}},\ \bibinfo {pages} {3}
  (\bibinfo {year} {2017})}\BibitemShut {NoStop}%
\bibitem [{\citenamefont {Shulman}\ \emph {et~al.}(2012)\citenamefont
  {Shulman}, \citenamefont {Dial}, \citenamefont {Harvey}, \citenamefont
  {Bluhm}, \citenamefont {Umansky},\ and\ \citenamefont
  {Yacoby}}]{Shulman_2012}%
  \BibitemOpen
  \bibfield  {author} {\bibinfo {author} {\bibfnamefont {M.~D.}\ \bibnamefont
  {Shulman}}, \bibinfo {author} {\bibfnamefont {O.~E.}\ \bibnamefont {Dial}},
  \bibinfo {author} {\bibfnamefont {S.~P.}\ \bibnamefont {Harvey}}, \bibinfo
  {author} {\bibfnamefont {H.}~\bibnamefont {Bluhm}}, \bibinfo {author}
  {\bibfnamefont {V.}~\bibnamefont {Umansky}}, \ and\ \bibinfo {author}
  {\bibfnamefont {A.}~\bibnamefont {Yacoby}},\ }\href {\doibase
  10.1126/science.1217692} {\bibfield  {journal} {\bibinfo  {journal}
  {Science}\ }\textbf {\bibinfo {volume} {336}},\ \bibinfo {pages} {202}
  (\bibinfo {year} {2012})}\BibitemShut {NoStop}%
\bibitem [{\citenamefont {Yoneda}\ \emph {et~al.}(2018)\citenamefont {Yoneda},
  \citenamefont {Takeda}, \citenamefont {Otsuka}, \citenamefont {Nakajima},
  \citenamefont {Delbecq}, \citenamefont {Allison}, \citenamefont {Honda},
  \citenamefont {Kodera}, \citenamefont {Oda}, \citenamefont {Hoshi},
  \citenamefont {Usami}, \citenamefont {Itoh},\ and\ \citenamefont
  {Tarucha}}]{Yoneda_2018}%
  \BibitemOpen
  \bibfield  {author} {\bibinfo {author} {\bibfnamefont {J.}~\bibnamefont
  {Yoneda}}, \bibinfo {author} {\bibfnamefont {K.}~\bibnamefont {Takeda}},
  \bibinfo {author} {\bibfnamefont {T.}~\bibnamefont {Otsuka}}, \bibinfo
  {author} {\bibfnamefont {T.}~\bibnamefont {Nakajima}}, \bibinfo {author}
  {\bibfnamefont {M.~R.}\ \bibnamefont {Delbecq}}, \bibinfo {author}
  {\bibfnamefont {G.}~\bibnamefont {Allison}}, \bibinfo {author} {\bibfnamefont
  {T.}~\bibnamefont {Honda}}, \bibinfo {author} {\bibfnamefont
  {T.}~\bibnamefont {Kodera}}, \bibinfo {author} {\bibfnamefont
  {S.}~\bibnamefont {Oda}}, \bibinfo {author} {\bibfnamefont {Y.}~\bibnamefont
  {Hoshi}}, \bibinfo {author} {\bibfnamefont {N.}~\bibnamefont {Usami}},
  \bibinfo {author} {\bibfnamefont {K.~M.}\ \bibnamefont {Itoh}}, \ and\
  \bibinfo {author} {\bibfnamefont {S.}~\bibnamefont {Tarucha}},\ }\href
  {\doibase 10.1038/s41565-017-0014-x} {\bibfield  {journal} {\bibinfo
  {journal} {Nature Nanotechnology}\ }\textbf {\bibinfo {volume} {13}},\
  \bibinfo {pages} {102} (\bibinfo {year} {2018})}\BibitemShut {NoStop}%
\bibitem [{\citenamefont {Onac}\ \emph {et~al.}(2006)\citenamefont {Onac},
  \citenamefont {Balestro}, \citenamefont {van Beveren}, \citenamefont
  {Hartmann}, \citenamefont {Nazarov},\ and\ \citenamefont
  {Kouwenhoven}}]{Onac_2006}%
  \BibitemOpen
  \bibfield  {author} {\bibinfo {author} {\bibfnamefont {E.}~\bibnamefont
  {Onac}}, \bibinfo {author} {\bibfnamefont {F.}~\bibnamefont {Balestro}},
  \bibinfo {author} {\bibfnamefont {L.~H.~W.}\ \bibnamefont {van Beveren}},
  \bibinfo {author} {\bibfnamefont {U.}~\bibnamefont {Hartmann}}, \bibinfo
  {author} {\bibfnamefont {Y.~V.}\ \bibnamefont {Nazarov}}, \ and\ \bibinfo
  {author} {\bibfnamefont {L.~P.}\ \bibnamefont {Kouwenhoven}},\ }\href
  {\doibase 10.1103/PhysRevLett.96.176601} {\bibfield  {journal} {\bibinfo
  {journal} {Phys. Rev. Lett.}\ }\textbf {\bibinfo {volume} {96}},\ \bibinfo
  {pages} {176601} (\bibinfo {year} {2006})}\BibitemShut {NoStop}%
\bibitem [{\citenamefont {Khrapai}\ \emph {et~al.}(2006)\citenamefont
  {Khrapai}, \citenamefont {Ludwig}, \citenamefont {Kotthaus}, \citenamefont
  {Tranitz},\ and\ \citenamefont {Wegscheider}}]{Khrapai_2006}%
  \BibitemOpen
  \bibfield  {author} {\bibinfo {author} {\bibfnamefont {V.~S.}\ \bibnamefont
  {Khrapai}}, \bibinfo {author} {\bibfnamefont {S.}~\bibnamefont {Ludwig}},
  \bibinfo {author} {\bibfnamefont {J.~P.}\ \bibnamefont {Kotthaus}}, \bibinfo
  {author} {\bibfnamefont {H.~P.}\ \bibnamefont {Tranitz}}, \ and\ \bibinfo
  {author} {\bibfnamefont {W.}~\bibnamefont {Wegscheider}},\ }\href {\doibase
  10.1103/PhysRevLett.97.176803} {\bibfield  {journal} {\bibinfo  {journal}
  {Phys. Rev. Lett.}\ }\textbf {\bibinfo {volume} {97}},\ \bibinfo {pages}
  {176803} (\bibinfo {year} {2006})}\BibitemShut {NoStop}%
\bibitem [{\citenamefont {Gustavsson}\ \emph {et~al.}(2007)\citenamefont
  {Gustavsson}, \citenamefont {Studer}, \citenamefont {Leturcq}, \citenamefont
  {Ihn}, \citenamefont {Ensslin}, \citenamefont {Driscoll},\ and\ \citenamefont
  {Gossard}}]{Gustavsson_2007}%
  \BibitemOpen
  \bibfield  {author} {\bibinfo {author} {\bibfnamefont {S.}~\bibnamefont
  {Gustavsson}}, \bibinfo {author} {\bibfnamefont {M.}~\bibnamefont {Studer}},
  \bibinfo {author} {\bibfnamefont {R.}~\bibnamefont {Leturcq}}, \bibinfo
  {author} {\bibfnamefont {T.}~\bibnamefont {Ihn}}, \bibinfo {author}
  {\bibfnamefont {K.}~\bibnamefont {Ensslin}}, \bibinfo {author} {\bibfnamefont
  {D.~C.}\ \bibnamefont {Driscoll}}, \ and\ \bibinfo {author} {\bibfnamefont
  {A.~C.}\ \bibnamefont {Gossard}},\ }\href {\doibase
  10.1103/PhysRevLett.99.206804} {\bibfield  {journal} {\bibinfo  {journal}
  {Phys. Rev. Lett.}\ }\textbf {\bibinfo {volume} {99}},\ \bibinfo {pages}
  {206804} (\bibinfo {year} {2007})}\BibitemShut {NoStop}%
\bibitem [{\citenamefont {Horibe}\ \emph {et~al.}(2015)\citenamefont {Horibe},
  \citenamefont {Kodera},\ and\ \citenamefont {Oda}}]{Horibe_2015}%
  \BibitemOpen
  \bibfield  {author} {\bibinfo {author} {\bibfnamefont {K.}~\bibnamefont
  {Horibe}}, \bibinfo {author} {\bibfnamefont {T.}~\bibnamefont {Kodera}}, \
  and\ \bibinfo {author} {\bibfnamefont {S.}~\bibnamefont {Oda}},\ }\href
  {\doibase 10.1063/1.4907894} {\bibfield  {journal} {\bibinfo  {journal}
  {Applied Physics Letters}\ }\textbf {\bibinfo {volume} {106}},\ \bibinfo
  {pages} {053119} (\bibinfo {year} {2015})}\BibitemShut {NoStop}%
\bibitem [{\citenamefont {Harvey-Collard}\ \emph {et~al.}(2017)\citenamefont
  {Harvey-Collard}, \citenamefont {Jacobson}, \citenamefont {Rudolph},
  \citenamefont {Dominguez}, \citenamefont {Ten~Eyck}, \citenamefont {Wendt},
  \citenamefont {Pluym}, \citenamefont {Gamble}, \citenamefont {Lilly},
  \citenamefont {Pioro-Ladri{\`e}re},\ and\ \citenamefont
  {Carroll}}]{Harvey-Collard_2017qubit}%
  \BibitemOpen
  \bibfield  {author} {\bibinfo {author} {\bibfnamefont {P.}~\bibnamefont
  {Harvey-Collard}}, \bibinfo {author} {\bibfnamefont {N.~T.}\ \bibnamefont
  {Jacobson}}, \bibinfo {author} {\bibfnamefont {M.}~\bibnamefont {Rudolph}},
  \bibinfo {author} {\bibfnamefont {J.}~\bibnamefont {Dominguez}}, \bibinfo
  {author} {\bibfnamefont {G.~A.}\ \bibnamefont {Ten~Eyck}}, \bibinfo {author}
  {\bibfnamefont {J.~R.}\ \bibnamefont {Wendt}}, \bibinfo {author}
  {\bibfnamefont {T.}~\bibnamefont {Pluym}}, \bibinfo {author} {\bibfnamefont
  {J.~K.}\ \bibnamefont {Gamble}}, \bibinfo {author} {\bibfnamefont {M.~P.}\
  \bibnamefont {Lilly}}, \bibinfo {author} {\bibfnamefont {M.}~\bibnamefont
  {Pioro-Ladri{\`e}re}}, \ and\ \bibinfo {author} {\bibfnamefont {M.~S.}\
  \bibnamefont {Carroll}},\ }\href {\doibase 10.1038/s41467-017-01113-2}
  {\bibfield  {journal} {\bibinfo  {journal} {Nature Communications}\ }\textbf
  {\bibinfo {volume} {8}},\ \bibinfo {pages} {1029} (\bibinfo {year}
  {2017})}\BibitemShut {NoStop}%
\bibitem [{\citenamefont {Kalra}\ \emph {et~al.}(2016)\citenamefont {Kalra},
  \citenamefont {Laucht}, \citenamefont {Dehollain}, \citenamefont {Bar},
  \citenamefont {Freer}, \citenamefont {Simmons}, \citenamefont {Muhonen},\
  and\ \citenamefont {Morello}}]{Kalra_2016}%
  \BibitemOpen
  \bibfield  {author} {\bibinfo {author} {\bibfnamefont {R.}~\bibnamefont
  {Kalra}}, \bibinfo {author} {\bibfnamefont {A.}~\bibnamefont {Laucht}},
  \bibinfo {author} {\bibfnamefont {J.~P.}\ \bibnamefont {Dehollain}}, \bibinfo
  {author} {\bibfnamefont {D.}~\bibnamefont {Bar}}, \bibinfo {author}
  {\bibfnamefont {S.}~\bibnamefont {Freer}}, \bibinfo {author} {\bibfnamefont
  {S.}~\bibnamefont {Simmons}}, \bibinfo {author} {\bibfnamefont {J.~T.}\
  \bibnamefont {Muhonen}}, \ and\ \bibinfo {author} {\bibfnamefont
  {A.}~\bibnamefont {Morello}},\ }\href {\doibase 10.1063/1.4959153} {\bibfield
   {journal} {\bibinfo  {journal} {Review of Scientific Instruments}\ }\textbf
  {\bibinfo {volume} {87}},\ \bibinfo {pages} {073905} (\bibinfo {year}
  {2016})}\BibitemShut {NoStop}%
\bibitem [{\citenamefont {Schoelkopf}\ \emph {et~al.}(1998)\citenamefont
  {Schoelkopf}, \citenamefont {Wahlgren}, \citenamefont {Kozhevnikov},
  \citenamefont {Delsing},\ and\ \citenamefont {Prober}}]{Schoelkopf_1998}%
  \BibitemOpen
  \bibfield  {author} {\bibinfo {author} {\bibfnamefont {R.~J.}\ \bibnamefont
  {Schoelkopf}}, \bibinfo {author} {\bibfnamefont {P.}~\bibnamefont
  {Wahlgren}}, \bibinfo {author} {\bibfnamefont {A.~A.}\ \bibnamefont
  {Kozhevnikov}}, \bibinfo {author} {\bibfnamefont {P.}~\bibnamefont
  {Delsing}}, \ and\ \bibinfo {author} {\bibfnamefont {D.~E.}\ \bibnamefont
  {Prober}},\ }\href {\doibase 10.1126/science.280.5367.1238} {\bibfield
  {journal} {\bibinfo  {journal} {Science}\ }\textbf {\bibinfo {volume}
  {280}},\ \bibinfo {pages} {1238} (\bibinfo {year} {1998})}\BibitemShut
  {NoStop}%
\bibitem [{\citenamefont {Aassime}\ \emph {et~al.}(2001)\citenamefont
  {Aassime}, \citenamefont {Johansson}, \citenamefont {Wendin}, \citenamefont
  {Schoelkopf},\ and\ \citenamefont {Delsing}}]{Aassime_2001}%
  \BibitemOpen
  \bibfield  {author} {\bibinfo {author} {\bibfnamefont {A.}~\bibnamefont
  {Aassime}}, \bibinfo {author} {\bibfnamefont {G.}~\bibnamefont {Johansson}},
  \bibinfo {author} {\bibfnamefont {G.}~\bibnamefont {Wendin}}, \bibinfo
  {author} {\bibfnamefont {R.~J.}\ \bibnamefont {Schoelkopf}}, \ and\ \bibinfo
  {author} {\bibfnamefont {P.}~\bibnamefont {Delsing}},\ }\href {\doibase
  10.1103/PhysRevLett.86.3376} {\bibfield  {journal} {\bibinfo  {journal}
  {Phys. Rev. Lett.}\ }\textbf {\bibinfo {volume} {86}},\ \bibinfo {pages}
  {3376} (\bibinfo {year} {2001})}\BibitemShut {NoStop}%
\bibitem [{\citenamefont {Reilly}\ \emph {et~al.}(2007)\citenamefont {Reilly},
  \citenamefont {Marcus}, \citenamefont {Hanson},\ and\ \citenamefont
  {Gossard}}]{Reilly_2007}%
  \BibitemOpen
  \bibfield  {author} {\bibinfo {author} {\bibfnamefont {D.~J.}\ \bibnamefont
  {Reilly}}, \bibinfo {author} {\bibfnamefont {C.~M.}\ \bibnamefont {Marcus}},
  \bibinfo {author} {\bibfnamefont {M.~P.}\ \bibnamefont {Hanson}}, \ and\
  \bibinfo {author} {\bibfnamefont {A.~C.}\ \bibnamefont {Gossard}},\ }\href
  {\doibase 10.1063/1.2794995} {\bibfield  {journal} {\bibinfo  {journal}
  {Applied Physics Letters}\ }\textbf {\bibinfo {volume} {91}},\ \bibinfo
  {pages} {162101} (\bibinfo {year} {2007})}\BibitemShut {NoStop}%
\bibitem [{\citenamefont {Barthel}\ \emph {et~al.}(2009)\citenamefont
  {Barthel}, \citenamefont {Reilly}, \citenamefont {Marcus}, \citenamefont
  {Hanson},\ and\ \citenamefont {Gossard}}]{Barthel_2009}%
  \BibitemOpen
  \bibfield  {author} {\bibinfo {author} {\bibfnamefont {C.}~\bibnamefont
  {Barthel}}, \bibinfo {author} {\bibfnamefont {D.~J.}\ \bibnamefont {Reilly}},
  \bibinfo {author} {\bibfnamefont {C.~M.}\ \bibnamefont {Marcus}}, \bibinfo
  {author} {\bibfnamefont {M.~P.}\ \bibnamefont {Hanson}}, \ and\ \bibinfo
  {author} {\bibfnamefont {A.~C.}\ \bibnamefont {Gossard}},\ }\href {\doibase
  10.1103/PhysRevLett.103.160503} {\bibfield  {journal} {\bibinfo  {journal}
  {Phys. Rev. Lett.}\ }\textbf {\bibinfo {volume} {103}},\ \bibinfo {pages}
  {160503} (\bibinfo {year} {2009})}\BibitemShut {NoStop}%
\bibitem [{\citenamefont {Barthel}\ \emph {et~al.}(2010)\citenamefont
  {Barthel}, \citenamefont {Kj{\ae}rgaard}, \citenamefont {Medford},
  \citenamefont {Stopa}, \citenamefont {Marcus}, \citenamefont {Hanson},\ and\
  \citenamefont {Gossard}}]{Barthel_2010}%
  \BibitemOpen
  \bibfield  {author} {\bibinfo {author} {\bibfnamefont {C.}~\bibnamefont
  {Barthel}}, \bibinfo {author} {\bibfnamefont {M.}~\bibnamefont
  {Kj{\ae}rgaard}}, \bibinfo {author} {\bibfnamefont {J.}~\bibnamefont
  {Medford}}, \bibinfo {author} {\bibfnamefont {M.}~\bibnamefont {Stopa}},
  \bibinfo {author} {\bibfnamefont {C.~M.}\ \bibnamefont {Marcus}}, \bibinfo
  {author} {\bibfnamefont {M.~P.}\ \bibnamefont {Hanson}}, \ and\ \bibinfo
  {author} {\bibfnamefont {A.~C.}\ \bibnamefont {Gossard}},\ }\href
  {http://dx.doi.org/10.1103/PhysRevB.81.161308} {\bibfield  {journal}
  {\bibinfo  {journal} {Physical Review B}\ }\textbf {\bibinfo {volume} {81}}
  (\bibinfo {year} {2010})}\BibitemShut {NoStop}%
\bibitem [{\citenamefont {Mason}\ \emph {et~al.}(2010)\citenamefont {Mason},
  \citenamefont {A.~Studenikin}, \citenamefont {Djurkovic}, \citenamefont
  {Sachrajda}, \citenamefont {Kycia}, \citenamefont {Gaudreau}, \citenamefont
  {Studenikin},\ and\ \citenamefont {Patricia~Kam}}]{Mason_2010}%
  \BibitemOpen
  \bibfield  {author} {\bibinfo {author} {\bibfnamefont {J.}~\bibnamefont
  {Mason}}, \bibinfo {author} {\bibfnamefont {S.}~\bibnamefont
  {A.~Studenikin}}, \bibinfo {author} {\bibfnamefont {B.}~\bibnamefont
  {Djurkovic}}, \bibinfo {author} {\bibfnamefont {A.}~\bibnamefont
  {Sachrajda}}, \bibinfo {author} {\bibfnamefont {J.}~\bibnamefont {Kycia}},
  \bibinfo {author} {\bibfnamefont {L.}~\bibnamefont {Gaudreau}}, \bibinfo
  {author} {\bibfnamefont {S.}~\bibnamefont {Studenikin}}, \ and\ \bibinfo
  {author} {\bibfnamefont {A.}~\bibnamefont {Patricia~Kam}},\ }\href
  {http://dx.doi.org/10.1016/j.physe.2009.11.108} {\bibfield  {journal}
  {\bibinfo  {journal} {Physica E: Low-dimensional Systems and Nanostructures}\
  }\textbf {\bibinfo {volume} {42}} (\bibinfo {year} {2010})}\BibitemShut
  {NoStop}%
\bibitem [{\citenamefont {Yuan}\ \emph {et~al.}(2012)\citenamefont {Yuan},
  \citenamefont {Yang}, \citenamefont {Savage}, \citenamefont {Lagally},
  \citenamefont {Eriksson},\ and\ \citenamefont {Rimberg}}]{Yuan_2012}%
  \BibitemOpen
  \bibfield  {author} {\bibinfo {author} {\bibfnamefont {M.}~\bibnamefont
  {Yuan}}, \bibinfo {author} {\bibfnamefont {Z.}~\bibnamefont {Yang}}, \bibinfo
  {author} {\bibfnamefont {D.~E.}\ \bibnamefont {Savage}}, \bibinfo {author}
  {\bibfnamefont {M.~G.}\ \bibnamefont {Lagally}}, \bibinfo {author}
  {\bibfnamefont {M.~A.}\ \bibnamefont {Eriksson}}, \ and\ \bibinfo {author}
  {\bibfnamefont {A.~J.}\ \bibnamefont {Rimberg}},\ }\href {\doibase
  10.1063/1.4754827} {\bibfield  {journal} {\bibinfo  {journal} {Applied
  Physics Letters}\ }\textbf {\bibinfo {volume} {101}},\ \bibinfo {pages}
  {142103} (\bibinfo {year} {2012})}\BibitemShut {NoStop}%
\bibitem [{\citenamefont {Verduijn}\ \emph {et~al.}(2014)\citenamefont
  {Verduijn}, \citenamefont {Vinet},\ and\ \citenamefont
  {Rogge}}]{Verduijn_2014}%
  \BibitemOpen
  \bibfield  {author} {\bibinfo {author} {\bibfnamefont {J.}~\bibnamefont
  {Verduijn}}, \bibinfo {author} {\bibfnamefont {M.}~\bibnamefont {Vinet}}, \
  and\ \bibinfo {author} {\bibfnamefont {S.}~\bibnamefont {Rogge}},\ }\href
  {\doibase 10.1063/1.4868423} {\bibfield  {journal} {\bibinfo  {journal}
  {Applied Physics Letters}\ }\textbf {\bibinfo {volume} {104}},\ \bibinfo
  {pages} {102107} (\bibinfo {year} {2014})}\BibitemShut {NoStop}%
\bibitem [{\citenamefont {Colless}\ \emph {et~al.}(2013)\citenamefont
  {Colless}, \citenamefont {Mahoney}, \citenamefont {Hornibrook}, \citenamefont
  {Doherty}, \citenamefont {Lu}, \citenamefont {Gossard},\ and\ \citenamefont
  {Reilly}}]{Colless_2013}%
  \BibitemOpen
  \bibfield  {author} {\bibinfo {author} {\bibfnamefont {J.~I.}\ \bibnamefont
  {Colless}}, \bibinfo {author} {\bibfnamefont {A.~C.}\ \bibnamefont
  {Mahoney}}, \bibinfo {author} {\bibfnamefont {J.~M.}\ \bibnamefont
  {Hornibrook}}, \bibinfo {author} {\bibfnamefont {A.~C.}\ \bibnamefont
  {Doherty}}, \bibinfo {author} {\bibfnamefont {H.}~\bibnamefont {Lu}},
  \bibinfo {author} {\bibfnamefont {A.~C.}\ \bibnamefont {Gossard}}, \ and\
  \bibinfo {author} {\bibfnamefont {D.~J.}\ \bibnamefont {Reilly}},\ }\href
  {http://dx.doi.org/10.1103/PhysRevLett.110.046805} {\bibfield  {journal}
  {\bibinfo  {journal} {Physical Review Letters}\ }\textbf {\bibinfo {volume}
  {110}} (\bibinfo {year} {2013})}\BibitemShut {NoStop}%
\bibitem [{\citenamefont {Stehlik}\ \emph {et~al.}(2015)\citenamefont
  {Stehlik}, \citenamefont {Liu}, \citenamefont {Quintana}, \citenamefont
  {Eichler}, \citenamefont {Hartke},\ and\ \citenamefont
  {Petta}}]{Stehlik_2015}%
  \BibitemOpen
  \bibfield  {author} {\bibinfo {author} {\bibfnamefont {J.}~\bibnamefont
  {Stehlik}}, \bibinfo {author} {\bibfnamefont {Y.-Y.}\ \bibnamefont {Liu}},
  \bibinfo {author} {\bibfnamefont {C.~M.}\ \bibnamefont {Quintana}}, \bibinfo
  {author} {\bibfnamefont {C.}~\bibnamefont {Eichler}}, \bibinfo {author}
  {\bibfnamefont {T.~R.}\ \bibnamefont {Hartke}}, \ and\ \bibinfo {author}
  {\bibfnamefont {J.~R.}\ \bibnamefont {Petta}},\ }\href {\doibase
  10.1103/PhysRevApplied.4.014018} {\bibfield  {journal} {\bibinfo  {journal}
  {Phys. Rev. Applied}\ }\textbf {\bibinfo {volume} {4}},\ \bibinfo {pages}
  {014018} (\bibinfo {year} {2015})}\BibitemShut {NoStop}%
\bibitem [{\citenamefont {Visscher}\ \emph {et~al.}(1996)\citenamefont
  {Visscher}, \citenamefont {Lindeman}, \citenamefont {Verbrugh}, \citenamefont
  {Hadley}, \citenamefont {Mooij},\ and\ \citenamefont {van~der
  Vleuten}}]{Visscher_1996}%
  \BibitemOpen
  \bibfield  {author} {\bibinfo {author} {\bibfnamefont {E.~H.}\ \bibnamefont
  {Visscher}}, \bibinfo {author} {\bibfnamefont {J.}~\bibnamefont {Lindeman}},
  \bibinfo {author} {\bibfnamefont {S.~M.}\ \bibnamefont {Verbrugh}}, \bibinfo
  {author} {\bibfnamefont {P.}~\bibnamefont {Hadley}}, \bibinfo {author}
  {\bibfnamefont {J.~E.}\ \bibnamefont {Mooij}}, \ and\ \bibinfo {author}
  {\bibfnamefont {W.}~\bibnamefont {van~der Vleuten}},\ }\href {\doibase
  10.1063/1.115622} {\bibfield  {journal} {\bibinfo  {journal} {Applied Physics
  Letters}\ }\textbf {\bibinfo {volume} {68}},\ \bibinfo {pages} {2014}
  (\bibinfo {year} {1996})}\BibitemShut {NoStop}%
\bibitem [{\citenamefont {Pettersson}\ \emph {et~al.}(1996)\citenamefont
  {Pettersson}, \citenamefont {Wahlgren}, \citenamefont {Delsing},
  \citenamefont {Haviland}, \citenamefont {Claeson}, \citenamefont {Rorsman},\
  and\ \citenamefont {Zirath}}]{Pettersson_1996}%
  \BibitemOpen
  \bibfield  {author} {\bibinfo {author} {\bibfnamefont {J.}~\bibnamefont
  {Pettersson}}, \bibinfo {author} {\bibfnamefont {P.}~\bibnamefont
  {Wahlgren}}, \bibinfo {author} {\bibfnamefont {P.}~\bibnamefont {Delsing}},
  \bibinfo {author} {\bibfnamefont {D.~B.}\ \bibnamefont {Haviland}}, \bibinfo
  {author} {\bibfnamefont {T.}~\bibnamefont {Claeson}}, \bibinfo {author}
  {\bibfnamefont {N.}~\bibnamefont {Rorsman}}, \ and\ \bibinfo {author}
  {\bibfnamefont {H.}~\bibnamefont {Zirath}},\ }\href {\doibase
  10.1103/PhysRevB.53.R13272} {\bibfield  {journal} {\bibinfo  {journal} {Phys.
  Rev. B}\ }\textbf {\bibinfo {volume} {53}},\ \bibinfo {pages} {R13272}
  (\bibinfo {year} {1996})}\BibitemShut {NoStop}%
\bibitem [{\citenamefont {Vink}\ \emph {et~al.}(2007)\citenamefont {Vink},
  \citenamefont {Nooitgedagt}, \citenamefont {Schouten}, \citenamefont
  {Vandersypen},\ and\ \citenamefont {Wegscheider}}]{Vink_2007}%
  \BibitemOpen
  \bibfield  {author} {\bibinfo {author} {\bibfnamefont {I.~T.}\ \bibnamefont
  {Vink}}, \bibinfo {author} {\bibfnamefont {T.}~\bibnamefont {Nooitgedagt}},
  \bibinfo {author} {\bibfnamefont {R.~N.}\ \bibnamefont {Schouten}}, \bibinfo
  {author} {\bibfnamefont {L.~M.~K.}\ \bibnamefont {Vandersypen}}, \ and\
  \bibinfo {author} {\bibfnamefont {W.}~\bibnamefont {Wegscheider}},\ }\href
  {\doibase 10.1063/1.2783265} {\bibfield  {journal} {\bibinfo  {journal}
  {Applied Physics Letters}\ }\textbf {\bibinfo {volume} {91}},\ \bibinfo
  {pages} {123512} (\bibinfo {year} {2007})}\BibitemShut {NoStop}%
\bibitem [{\citenamefont {Curry}\ \emph {et~al.}(2015)\citenamefont {Curry},
  \citenamefont {England}, \citenamefont {Bishop}, \citenamefont {Ten-Eyck},
  \citenamefont {Wendt}, \citenamefont {Pluym}, \citenamefont {Lilly},
  \citenamefont {Carr},\ and\ \citenamefont {Carroll}}]{Curry_2015}%
  \BibitemOpen
  \bibfield  {author} {\bibinfo {author} {\bibfnamefont {M.~J.}\ \bibnamefont
  {Curry}}, \bibinfo {author} {\bibfnamefont {T.~D.}\ \bibnamefont {England}},
  \bibinfo {author} {\bibfnamefont {N.~C.}\ \bibnamefont {Bishop}}, \bibinfo
  {author} {\bibfnamefont {G.}~\bibnamefont {Ten-Eyck}}, \bibinfo {author}
  {\bibfnamefont {J.~R.}\ \bibnamefont {Wendt}}, \bibinfo {author}
  {\bibfnamefont {T.}~\bibnamefont {Pluym}}, \bibinfo {author} {\bibfnamefont
  {M.~P.}\ \bibnamefont {Lilly}}, \bibinfo {author} {\bibfnamefont {S.~M.}\
  \bibnamefont {Carr}}, \ and\ \bibinfo {author} {\bibfnamefont {M.~S.}\
  \bibnamefont {Carroll}},\ }\href {\doibase 10.1063/1.4921308} {\bibfield
  {journal} {\bibinfo  {journal} {Applied Physics Letters}\ }\textbf {\bibinfo
  {volume} {106}},\ \bibinfo {pages} {203505} (\bibinfo {year}
  {2015})}\BibitemShut {NoStop}%
\bibitem [{\citenamefont {Tracy}\ \emph {et~al.}(2016)\citenamefont {Tracy},
  \citenamefont {Luhman}, \citenamefont {Carr}, \citenamefont {Bishop},
  \citenamefont {Ten~Eyck}, \citenamefont {Pluym}, \citenamefont {Wendt},
  \citenamefont {Lilly},\ and\ \citenamefont {Carroll}}]{Tracy_2016}%
  \BibitemOpen
  \bibfield  {author} {\bibinfo {author} {\bibfnamefont {L.~A.}\ \bibnamefont
  {Tracy}}, \bibinfo {author} {\bibfnamefont {D.~R.}\ \bibnamefont {Luhman}},
  \bibinfo {author} {\bibfnamefont {S.~M.}\ \bibnamefont {Carr}}, \bibinfo
  {author} {\bibfnamefont {N.~C.}\ \bibnamefont {Bishop}}, \bibinfo {author}
  {\bibfnamefont {G.~A.}\ \bibnamefont {Ten~Eyck}}, \bibinfo {author}
  {\bibfnamefont {T.}~\bibnamefont {Pluym}}, \bibinfo {author} {\bibfnamefont
  {J.~R.}\ \bibnamefont {Wendt}}, \bibinfo {author} {\bibfnamefont {M.~P.}\
  \bibnamefont {Lilly}}, \ and\ \bibinfo {author} {\bibfnamefont {M.~S.}\
  \bibnamefont {Carroll}},\ }\href {\doibase 10.1063/1.4941421} {\bibfield
  {journal} {\bibinfo  {journal} {Applied Physics Letters}\ }\textbf {\bibinfo
  {volume} {108}},\ \bibinfo {pages} {063101} (\bibinfo {year}
  {2016})}\BibitemShut {NoStop}%
\bibitem [{\citenamefont {Joseph}\ \emph {et~al.}(1995)\citenamefont {Joseph},
  \citenamefont {Cressler},\ and\ \citenamefont {Richey}}]{Joseph_1995}%
  \BibitemOpen
  \bibfield  {author} {\bibinfo {author} {\bibfnamefont {A.~J.}\ \bibnamefont
  {Joseph}}, \bibinfo {author} {\bibfnamefont {J.~D.}\ \bibnamefont
  {Cressler}}, \ and\ \bibinfo {author} {\bibfnamefont {D.~M.}\ \bibnamefont
  {Richey}},\ }\href {\doibase 10.1109/55.790731} {\bibfield  {journal}
  {\bibinfo  {journal} {IEEE Electron Device Letters}\ }\textbf {\bibinfo
  {volume} {16}},\ \bibinfo {pages} {268} (\bibinfo {year} {1995})}\BibitemShut
  {NoStop}%
\bibitem [{\citenamefont {Najafizadeh}\ \emph {et~al.}(2009)\citenamefont
  {Najafizadeh}, \citenamefont {Adams}, \citenamefont {Phillips}, \citenamefont
  {Moen}, \citenamefont {Cressler}, \citenamefont {Aslam}, \citenamefont
  {Stevenson},\ and\ \citenamefont {Meloy}}]{Najafizedeh_2009}%
  \BibitemOpen
  \bibfield  {author} {\bibinfo {author} {\bibfnamefont {L.}~\bibnamefont
  {Najafizadeh}}, \bibinfo {author} {\bibfnamefont {J.~S.}\ \bibnamefont
  {Adams}}, \bibinfo {author} {\bibfnamefont {S.~D.}\ \bibnamefont {Phillips}},
  \bibinfo {author} {\bibfnamefont {K.~A.}\ \bibnamefont {Moen}}, \bibinfo
  {author} {\bibfnamefont {J.~D.}\ \bibnamefont {Cressler}}, \bibinfo {author}
  {\bibfnamefont {S.}~\bibnamefont {Aslam}}, \bibinfo {author} {\bibfnamefont
  {T.~R.}\ \bibnamefont {Stevenson}}, \ and\ \bibinfo {author} {\bibfnamefont
  {R.~M.}\ \bibnamefont {Meloy}},\ }\href {\doibase 10.1109/LED.2009.2016767}
  {\bibfield  {journal} {\bibinfo  {journal} {IEEE Electron Device Letters}\
  }\textbf {\bibinfo {volume} {30}},\ \bibinfo {pages} {508} (\bibinfo {year}
  {2009})}\BibitemShut {NoStop}%
\bibitem [{\citenamefont {Curry}\ \emph {et~al.}(2016)\citenamefont {Curry},
  \citenamefont {England}, \citenamefont {Wendt}, \citenamefont {Pluym},
  \citenamefont {Lilly}, \citenamefont {Carr},\ and\ \citenamefont
  {Carroll}}]{Curry_2016}%
  \BibitemOpen
  \bibfield  {author} {\bibinfo {author} {\bibfnamefont {M.}~\bibnamefont
  {Curry}}, \bibinfo {author} {\bibfnamefont {T.}~\bibnamefont {England}},
  \bibinfo {author} {\bibfnamefont {J.}~\bibnamefont {Wendt}}, \bibinfo
  {author} {\bibfnamefont {T.}~\bibnamefont {Pluym}}, \bibinfo {author}
  {\bibfnamefont {M.}~\bibnamefont {Lilly}}, \bibinfo {author} {\bibfnamefont
  {S.}~\bibnamefont {Carr}}, \ and\ \bibinfo {author} {\bibfnamefont
  {M.}~\bibnamefont {Carroll}},\ }\href
  {http://meetings.aps.org/Meeting/MAR16/Session/P45.11} {\bibfield  {journal}
  {\bibinfo  {journal} {Bulletin of the American Physical Society}\ }\textbf
  {\bibinfo {volume} {61}} (\bibinfo {year} {2016})}\BibitemShut {NoStop}%
\bibitem [{\citenamefont {Ying}\ \emph {et~al.}(2017)\citenamefont {Ying},
  \citenamefont {Wier}, \citenamefont {Dark}, \citenamefont {Lourenco},
  \citenamefont {Ge}, \citenamefont {Omprakash}, \citenamefont {Mourigal},
  \citenamefont {Davidovic},\ and\ \citenamefont {Cressler}}]{Ying_2017}%
  \BibitemOpen
  \bibfield  {author} {\bibinfo {author} {\bibfnamefont {H.}~\bibnamefont
  {Ying}}, \bibinfo {author} {\bibfnamefont {B.~R.}\ \bibnamefont {Wier}},
  \bibinfo {author} {\bibfnamefont {J.}~\bibnamefont {Dark}}, \bibinfo {author}
  {\bibfnamefont {N.~E.}\ \bibnamefont {Lourenco}}, \bibinfo {author}
  {\bibfnamefont {L.}~\bibnamefont {Ge}}, \bibinfo {author} {\bibfnamefont
  {A.~P.}\ \bibnamefont {Omprakash}}, \bibinfo {author} {\bibfnamefont
  {M.}~\bibnamefont {Mourigal}}, \bibinfo {author} {\bibfnamefont
  {D.}~\bibnamefont {Davidovic}}, \ and\ \bibinfo {author} {\bibfnamefont
  {J.~D.}\ \bibnamefont {Cressler}},\ }\href {\doibase
  10.1109/led.2016.2633465} {\bibfield  {journal} {\bibinfo  {journal} {IEEE
  Electron Device Letters}\ }\textbf {\bibinfo {volume} {38}},\ \bibinfo
  {pages} {12} (\bibinfo {year} {2017})}\BibitemShut {NoStop}%
\bibitem [{\citenamefont {Davidovi\ifmmode~\acute{c}\else \'{c}\fi{}}\ \emph
  {et~al.}(2017)\citenamefont {Davidovi\ifmmode~\acute{c}\else \'{c}\fi{}},
  \citenamefont {Ying}, \citenamefont {Dark}, \citenamefont {Wier},
  \citenamefont {Ge}, \citenamefont {Lourenco}, \citenamefont {Omprakash},
  \citenamefont {Mourigal},\ and\ \citenamefont {Cressler}}]{Davidovic_2017}%
  \BibitemOpen
  \bibfield  {author} {\bibinfo {author} {\bibfnamefont {D.}~\bibnamefont
  {Davidovi\ifmmode~\acute{c}\else \'{c}\fi{}}}, \bibinfo {author}
  {\bibfnamefont {H.}~\bibnamefont {Ying}}, \bibinfo {author} {\bibfnamefont
  {J.}~\bibnamefont {Dark}}, \bibinfo {author} {\bibfnamefont {B.~R.}\
  \bibnamefont {Wier}}, \bibinfo {author} {\bibfnamefont {L.}~\bibnamefont
  {Ge}}, \bibinfo {author} {\bibfnamefont {N.~E.}\ \bibnamefont {Lourenco}},
  \bibinfo {author} {\bibfnamefont {A.~P.}\ \bibnamefont {Omprakash}}, \bibinfo
  {author} {\bibfnamefont {M.}~\bibnamefont {Mourigal}}, \ and\ \bibinfo
  {author} {\bibfnamefont {J.~D.}\ \bibnamefont {Cressler}},\ }\href {\doibase
  10.1103/PhysRevApplied.8.024015} {\bibfield  {journal} {\bibinfo  {journal}
  {Phys. Rev. Applied}\ }\textbf {\bibinfo {volume} {8}},\ \bibinfo {pages}
  {024015} (\bibinfo {year} {2017})}\BibitemShut {NoStop}%
\bibitem [{\citenamefont {Horowitz}\ \emph {et~al.}(1980)\citenamefont
  {Horowitz}, \citenamefont {Hill},\ and\ \citenamefont
  {Robinson}}]{Horowitz_1980}%
  \BibitemOpen
  \bibfield  {author} {\bibinfo {author} {\bibfnamefont {P.}~\bibnamefont
  {Horowitz}}, \bibinfo {author} {\bibfnamefont {W.}~\bibnamefont {Hill}}, \
  and\ \bibinfo {author} {\bibfnamefont {I.}~\bibnamefont {Robinson}},\
  }\href@noop {} {\emph {\bibinfo {title} {The Art of Electronics}}},\
  Vol.~\bibinfo {volume} {2}\ (\bibinfo  {publisher} {Cambridge University
  Press Cambridge},\ \bibinfo {year} {1980})\BibitemShut {NoStop}%
\bibitem [{\citenamefont {Jock}\ \emph {et~al.}(2018)\citenamefont {Jock},
  \citenamefont {Jacobson}, \citenamefont {Harvey-Collard}, \citenamefont
  {Mounce}, \citenamefont {Srinivasa}, \citenamefont {Ward}, \citenamefont
  {Anderson}, \citenamefont {Manginell}, \citenamefont {Wendt}, \citenamefont
  {Rudolph}, \citenamefont {Pluym}, \citenamefont {Gamble}, \citenamefont
  {Baczewski}, \citenamefont {Witzel},\ and\ \citenamefont
  {Carroll}}]{Jock_2018}%
  \BibitemOpen
  \bibfield  {author} {\bibinfo {author} {\bibfnamefont {R.~M.}\ \bibnamefont
  {Jock}}, \bibinfo {author} {\bibfnamefont {N.~T.}\ \bibnamefont {Jacobson}},
  \bibinfo {author} {\bibfnamefont {P.}~\bibnamefont {Harvey-Collard}},
  \bibinfo {author} {\bibfnamefont {A.~M.}\ \bibnamefont {Mounce}}, \bibinfo
  {author} {\bibfnamefont {V.}~\bibnamefont {Srinivasa}}, \bibinfo {author}
  {\bibfnamefont {D.~R.}\ \bibnamefont {Ward}}, \bibinfo {author}
  {\bibfnamefont {J.}~\bibnamefont {Anderson}}, \bibinfo {author}
  {\bibfnamefont {R.}~\bibnamefont {Manginell}}, \bibinfo {author}
  {\bibfnamefont {J.~R.}\ \bibnamefont {Wendt}}, \bibinfo {author}
  {\bibfnamefont {M.}~\bibnamefont {Rudolph}}, \bibinfo {author} {\bibfnamefont
  {T.}~\bibnamefont {Pluym}}, \bibinfo {author} {\bibfnamefont {J.~K.}\
  \bibnamefont {Gamble}}, \bibinfo {author} {\bibfnamefont {A.~D.}\
  \bibnamefont {Baczewski}}, \bibinfo {author} {\bibfnamefont {W.~M.}\
  \bibnamefont {Witzel}}, \ and\ \bibinfo {author} {\bibfnamefont {M.~S.}\
  \bibnamefont {Carroll}},\ }\href {\doibase 10.1038/s41467-018-04200-0}
  {\bibfield  {journal} {\bibinfo  {journal} {Nature Communications}\ }\textbf
  {\bibinfo {volume} {9}},\ \bibinfo {pages} {1768} (\bibinfo {year}
  {2018})}\BibitemShut {NoStop}%
\bibitem [{\citenamefont {England}\ \emph {et~al.}(2017)\citenamefont
  {England}, \citenamefont {Curry}, \citenamefont {Carr}, \citenamefont
  {Mounce}, \citenamefont {Jock}, \citenamefont {Sharma}, \citenamefont
  {Bureau-Oxton}, \citenamefont {Rudolph}, \citenamefont {Hardin},\ and\
  \citenamefont {Carroll}}]{England_2017}%
  \BibitemOpen
  \bibfield  {author} {\bibinfo {author} {\bibfnamefont {T.}~\bibnamefont
  {England}}, \bibinfo {author} {\bibfnamefont {M.}~\bibnamefont {Curry}},
  \bibinfo {author} {\bibfnamefont {S.}~\bibnamefont {Carr}}, \bibinfo {author}
  {\bibfnamefont {A.}~\bibnamefont {Mounce}}, \bibinfo {author} {\bibfnamefont
  {R.}~\bibnamefont {Jock}}, \bibinfo {author} {\bibfnamefont {P.}~\bibnamefont
  {Sharma}}, \bibinfo {author} {\bibfnamefont {C.}~\bibnamefont
  {Bureau-Oxton}}, \bibinfo {author} {\bibfnamefont {M.}~\bibnamefont
  {Rudolph}}, \bibinfo {author} {\bibfnamefont {T.}~\bibnamefont {Hardin}}, \
  and\ \bibinfo {author} {\bibfnamefont {M.}~\bibnamefont {Carroll}},\ }\href
  {http://meetings.aps.org/Meeting/MAR17/Event/291832} {\bibfield  {journal}
  {\bibinfo  {journal} {Bulletin of the American Physical Society}\ }\textbf
  {\bibinfo {volume} {62}} (\bibinfo {year} {2017})}\BibitemShut {NoStop}%
\bibitem [{\citenamefont {Tracy}\ \emph {et~al.}(2018)\citenamefont {Tracy},
  \citenamefont {Reno},\ and\ \citenamefont {Hargett}}]{Tracy_2018}%
  \BibitemOpen
  \bibfield  {author} {\bibinfo {author} {\bibfnamefont {L.}~\bibnamefont
  {Tracy}}, \bibinfo {author} {\bibfnamefont {J.}~\bibnamefont {Reno}}, \ and\
  \bibinfo {author} {\bibfnamefont {T.}~\bibnamefont {Hargett}},\ }\href
  {https://meetings.aps.org/Meeting/MAR18/Session/P28.6} {\bibfield  {journal}
  {\bibinfo  {journal} {Bulletin of the American Physical Society}\ } (\bibinfo
  {year} {2018})}\BibitemShut {NoStop}%
\bibitem [{\citenamefont {Nordberg}\ \emph {et~al.}(2009)\citenamefont
  {Nordberg}, \citenamefont {Stalford}, \citenamefont {Young}, \citenamefont
  {Ten~Eyck}, \citenamefont {Eng}, \citenamefont {Tracy}, \citenamefont
  {Childs}, \citenamefont {Wendt}, \citenamefont {Grubbs}, \citenamefont
  {Stevens}, \citenamefont {Lilly}, \citenamefont {Eriksson},\ and\
  \citenamefont {Carroll}}]{Nordberg_2009}%
  \BibitemOpen
  \bibfield  {author} {\bibinfo {author} {\bibfnamefont {E.~P.}\ \bibnamefont
  {Nordberg}}, \bibinfo {author} {\bibfnamefont {H.~L.}\ \bibnamefont
  {Stalford}}, \bibinfo {author} {\bibfnamefont {R.}~\bibnamefont {Young}},
  \bibinfo {author} {\bibfnamefont {G.~A.}\ \bibnamefont {Ten~Eyck}}, \bibinfo
  {author} {\bibfnamefont {K.}~\bibnamefont {Eng}}, \bibinfo {author}
  {\bibfnamefont {L.~A.}\ \bibnamefont {Tracy}}, \bibinfo {author}
  {\bibfnamefont {K.~D.}\ \bibnamefont {Childs}}, \bibinfo {author}
  {\bibfnamefont {J.~R.}\ \bibnamefont {Wendt}}, \bibinfo {author}
  {\bibfnamefont {R.~K.}\ \bibnamefont {Grubbs}}, \bibinfo {author}
  {\bibfnamefont {J.}~\bibnamefont {Stevens}}, \bibinfo {author} {\bibfnamefont
  {M.~P.}\ \bibnamefont {Lilly}}, \bibinfo {author} {\bibfnamefont {M.~A.}\
  \bibnamefont {Eriksson}}, \ and\ \bibinfo {author} {\bibfnamefont {M.~S.}\
  \bibnamefont {Carroll}},\ }\href {\doibase 10.1063/1.3259416} {\bibfield
  {journal} {\bibinfo  {journal} {Applied Physics Letters}\ }\textbf {\bibinfo
  {volume} {95}},\ \bibinfo {pages} {202102} (\bibinfo {year}
  {2009})}\BibitemShut {NoStop}%
\bibitem [{\citenamefont {Knapp}\ \emph {et~al.}(2018)\citenamefont {Knapp},
  \citenamefont {Dodson}, \citenamefont {Thorgrimsson}, \citenamefont {Savage},
  \citenamefont {Lagally}, \citenamefont {Coppersmith},\ and\ \citenamefont
  {Eriksson}}]{Knapp_2018}%
  \BibitemOpen
  \bibfield  {author} {\bibinfo {author} {\bibfnamefont {T.}~\bibnamefont
  {Knapp}}, \bibinfo {author} {\bibfnamefont {J.}~\bibnamefont {Dodson}},
  \bibinfo {author} {\bibfnamefont {B.}~\bibnamefont {Thorgrimsson}}, \bibinfo
  {author} {\bibfnamefont {D.}~\bibnamefont {Savage}}, \bibinfo {author}
  {\bibfnamefont {M.}~\bibnamefont {Lagally}}, \bibinfo {author} {\bibfnamefont
  {S.}~\bibnamefont {Coppersmith}}, \ and\ \bibinfo {author} {\bibfnamefont
  {M.}~\bibnamefont {Eriksson}},\ }\href
  {http://meetings.aps.org/Meeting/MAR18/Session/P28.5} {\bibfield  {journal}
  {\bibinfo  {journal} {Bulletin of the American Physical Society}\ }\textbf
  {\bibinfo {volume} {63}} (\bibinfo {year} {2018})}\BibitemShut {NoStop}%
\bibitem [{\citenamefont {Beenakker}\ and\ \citenamefont
  {Sch{\"o}nenberger}(2003)}]{Beenakker_2003}%
  \BibitemOpen
  \bibfield  {author} {\bibinfo {author} {\bibfnamefont {C.}~\bibnamefont
  {Beenakker}}\ and\ \bibinfo {author} {\bibfnamefont {C.}~\bibnamefont
  {Sch{\"o}nenberger}},\ }\href {\doibase 10.1063/1.1583532} {\bibfield
  {journal} {\bibinfo  {journal} {Physics Today}\ }\textbf {\bibinfo {volume}
  {56}},\ \bibinfo {pages} {37} (\bibinfo {year} {2003})}\BibitemShut {NoStop}%
\bibitem [{\citenamefont {Kafanov}\ and\ \citenamefont
  {Delsing}(2009)}]{Kafanov_2009}%
  \BibitemOpen
  \bibfield  {author} {\bibinfo {author} {\bibfnamefont {S.}~\bibnamefont
  {Kafanov}}\ and\ \bibinfo {author} {\bibfnamefont {P.}~\bibnamefont
  {Delsing}},\ }\href {\doibase 10.1103/PhysRevB.80.155320} {\bibfield
  {journal} {\bibinfo  {journal} {Phys. Rev. B}\ }\textbf {\bibinfo {volume}
  {80}},\ \bibinfo {pages} {155320} (\bibinfo {year} {2009})}\BibitemShut
  {NoStop}%
\end{thebibliography}

%

\appendix

\section{SET Geometries And Details\label{sec:SET_Geometries}}

The SET connected to the AC-HBT uses a single layer doped poly-Si
electrode structure on 50 nm thick SiO\textsubscript{2}, providing
a mobility of 19,500 cm\textsuperscript{2}/Vs at 4 K. The poly-Si
gate layer is etch-defined into electrodes that control the formation
of the SET (upper left in Figure \ref{fig:ac-hbt_circuit_diagram}(a)
SEM image) and two quantum dots (under gates RD and LD). Regions of
electron enhancement are indicated by the highlighted regions.

The Si-MOS device in the CB-HBT circuit is similar to the Si-MOS device
in the AC-HBT circuit with the exception that the SiO\textsubscript{2}
layer is 35 nm thick and the bottom layer is isotopically purified
silicon (500 ppm \textsuperscript{29}Si). The \textsuperscript{28}Si
isotope has no net nuclear spin, therefore it is ideal for qubits
to be formed in because decoherence due to magnetic noise is highly
suppressed. Phosphorous (\textsuperscript{31}P) donor atoms are imbedded
in the \textsuperscript{28}Si layer using ion implantation near where
the quantum dot is intended to be formed (red dot in Figure \ref{fig:cb-hbt_circuit_diagram}(a)
SEM image).

\begin{figure}
\begin{centering}
\includegraphics[width=8.5cm]{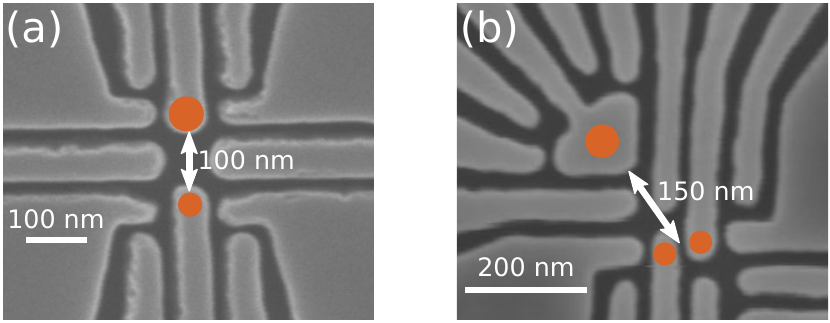}
\par\end{centering}
\caption{(a) SEM image of Si-MOS device used in CB-HBT circuit. The edge of
the SET (larger orange dot) is roughly 100 nm away from the quantum
dot (smaller orange dot). (b) SEM image of the Si-MOS device used
in the AC-HBT circuit. The proximity of the SET to the double quantum
dot system is 50\% further away at roughly 150 nm. }

\label{fig:sem_geometry_comparison}
\end{figure}

The CB-HBT and AC-HBT were characterized using different Si-MOS devices
possessing different electrostatic gate layouts (Figure \ref{fig:sem_geometry_comparison}).
The geometry of the gate layout affects the mutual capacitance between
the SET and the quantum dot. More capacitive coupling results in larger
changes in the electrochemical potential of the charge-sensor for
a given quantum dot charging event \citep{Nordberg_2009}. Since changes
in electrochemical potential of the charge-sensor result in changes
in current through the charge-sensor, larger changes result in larger
signal. Therefore, more mutual capacitance leads to larger readout
signals, faster readout times, and higher readout fidelity.

The gate geometry used in the Si-MOS device connected to the CB-HBT
had the SET 33\% closer to the quantum dot than in the Si-MOS device
connected to the AC-HBT. The closer SET proximity in the CB-HBT resulted
in an increase in sensitivity of approximately 34\%. We compare the
sensitivity of both circuits by dividing the voltage shift of the
dot occupancy transition by the charge-sensor Coulomb blockade peak
period. For the CB-HBT, the voltage shift was 18 mV and the charge-sensor
period was 337 mV (5.34\% change). For the AC-HBT, the voltage shift
was 12 mV and the charge-sensor period was 350 mV (4\% change). Therefore,
the SET in the CB-HBT was around 34\% more sensitive to charging events
than the AC-HBT.

\begin{figure}
\begin{centering}
\includegraphics[width=8.5cm]{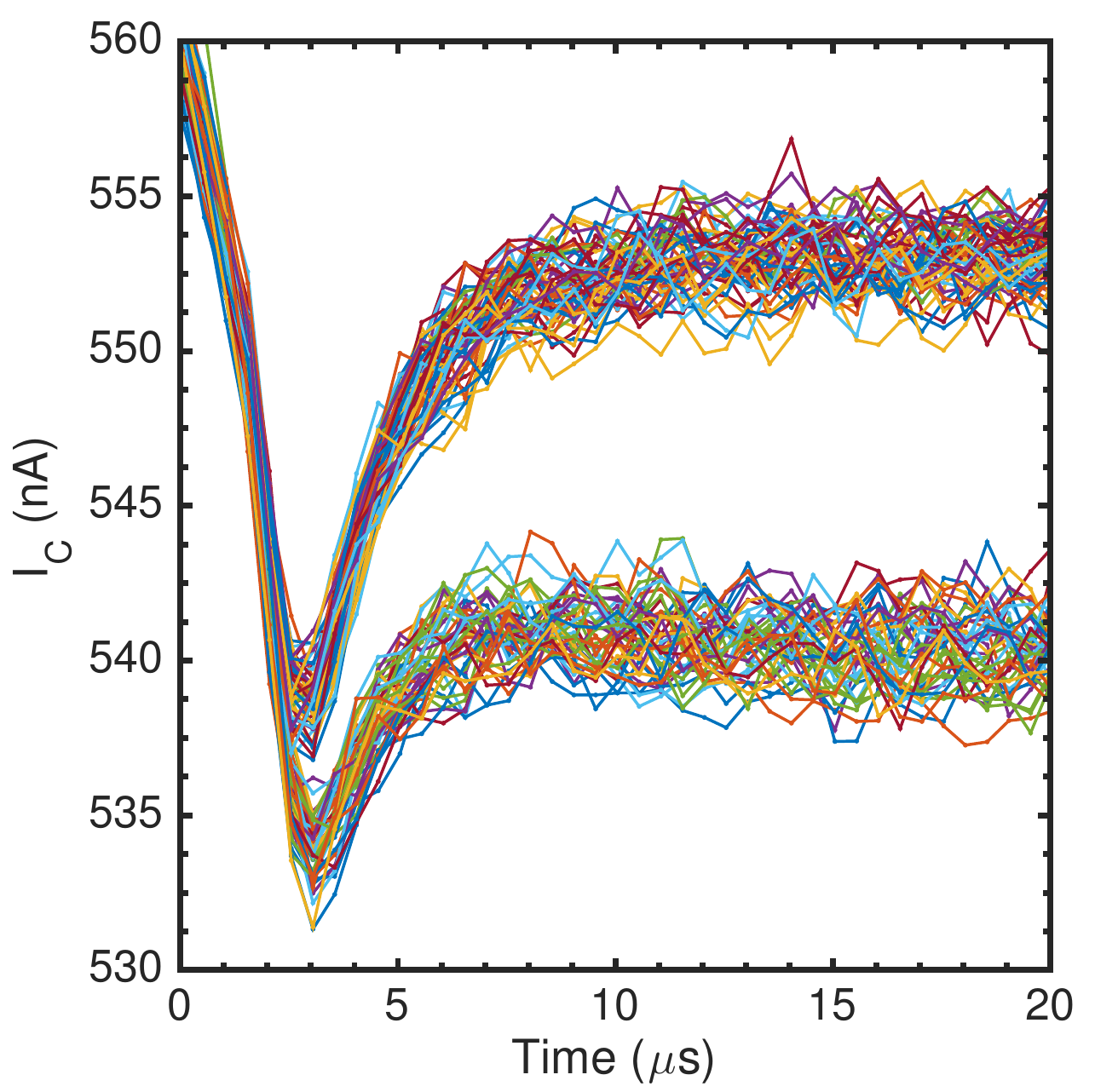}
\par\end{centering}
\caption{100 single-shot traces for the CB-HBT charge readout. Slower response
time is compensated for by larger signal separation at earlier times
relative to the AC-HBT readout.}
\end{figure}

\section{Current-Biasing Effect of CB-HBT Circuit\label{sec:CB-HBT_Biasing_Effect}}

\begin{figure}
\begin{centering}
\includegraphics[width=8.5cm]{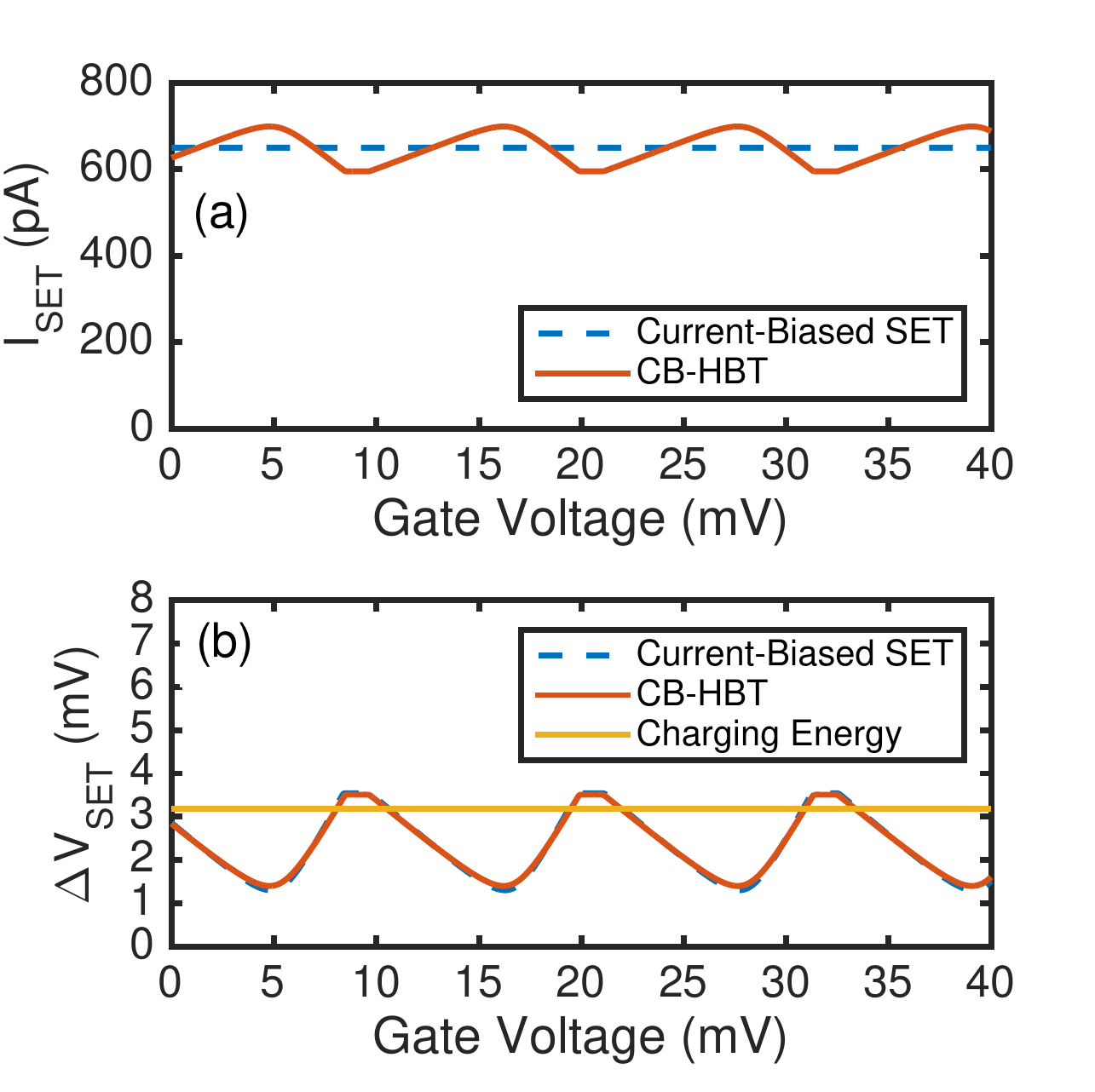}
\par\end{centering}
\caption{Comparison of CB-HBT and current-biased SET Verilog-A models. The
top plot shows the drain current of the SET as a function of gate
voltage. The current is modulated much less in this condition than
in a constant voltage-biased circuit. The bottom plot shows voltage
across the SET as a function of gate voltage. The overlap between
the two curves shows that the CB-HBT circuit is effectively equivalent
to current-biasing the SET.}

\label{fig:cb_effect_modeling}
\end{figure}

Since the node that connects the SET source to the HBT base is floating,
the bias across the SET cannot be set to a fixed voltage in the CB-HBT
circuit. Verilog-A models were created to simulate the behavior of
the circuit when biasing the SET through multiple regions of Coulomb
blockade via an electrostatic gate. As the SET resistance changes
due to Coulomb blockade, the source-drain bias across the SET changes
to allow current to flow into the base of the HBT (Figure \ref{fig:cb_effect_modeling}(b)).
In order for this to happen, the HBT trades base-emitter voltage for
minimal impact to operation. Although the trade in voltage results
in a relatively small change in HBT collector current during, for
example, a single-shot readout event, this signal is approximately
100 larger than the SET source-drain signal without an HBT (e.g. \textgreek{D}I\textsubscript{C}
= 10 nA vs. \textgreek{D}I\textsubscript{SET} = 100 pA).

\begin{figure}
\begin{centering}
\includegraphics[width=8.5cm]{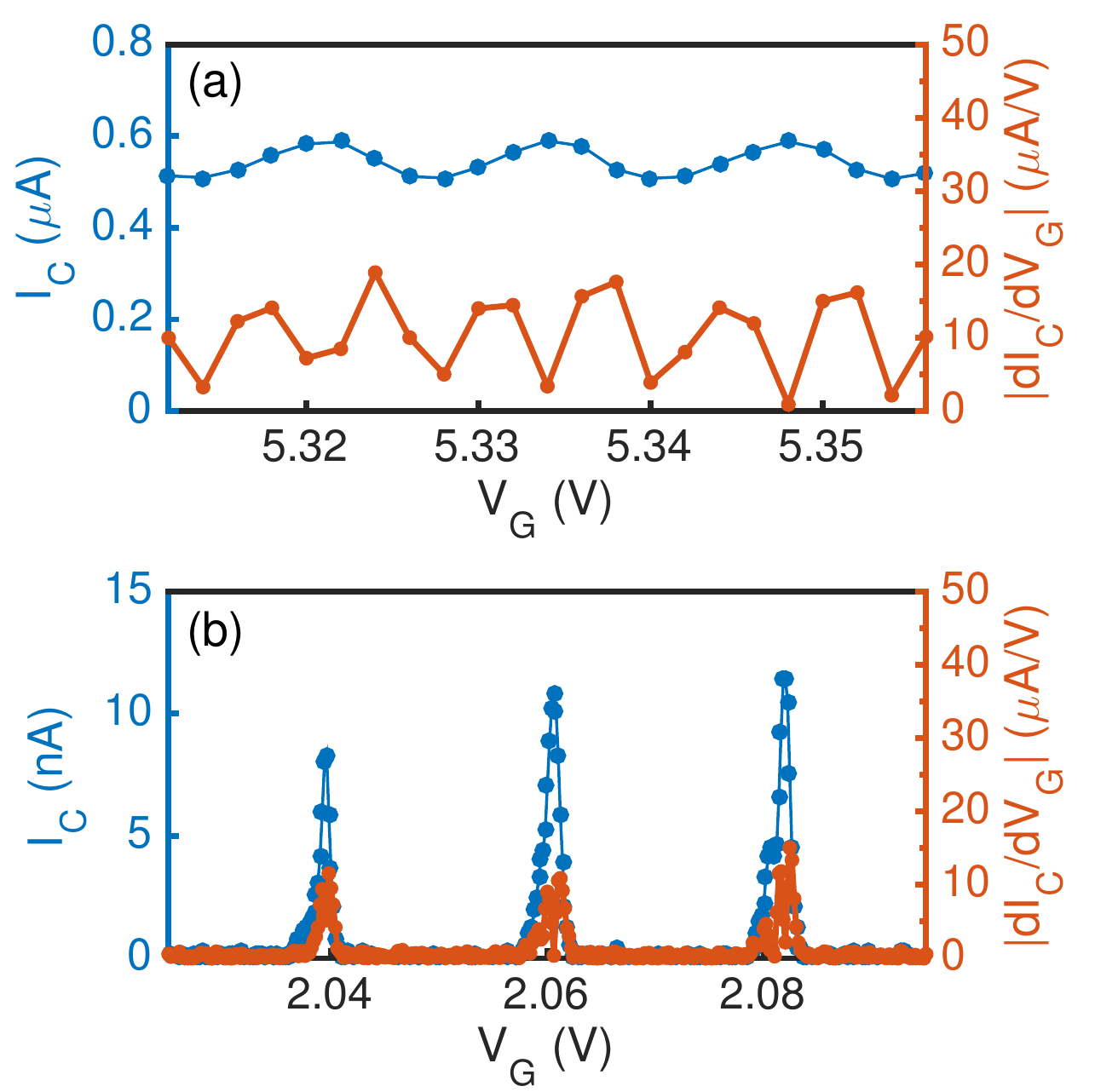}
\par\end{centering}
\caption{(a) Coulomb blockade peaks for the CB-HBT (blue). The absolute value
of the sensitivity is plotted as orange points. Since there is almost
always positive or negative blockade slope, the absolute value of
the sensitivity remains positive for most of the range plotted. (b)
Coulomb blockade peaks for the AC-HBT (blue). The absolute value of
the sensitivity is plotted as orange points as well.}

\label{fig:current-biasing_effect_of_CB-HBT}
\end{figure}

The Verilog-A model estimates the small signal resistances as: r\textsubscript{set}
= 200 k\textgreek{W} and r\textsubscript{\textgreek{p}} = 10 M\textgreek{W}
(where r\textsubscript{\textgreek{p}} is the small signal resistance
of the base-emitter junction). Most of the emitter bias voltage is
across the base-emitter junction at all times (since r\textsubscript{set}
<\textcompwordmark < r\textsubscript{\textgreek{p}}), therefore
the CB-HBT is a current-biasing circuit. The current-biasing behavior
is highlighted in Figure \ref{fig:current-biasing_effect_of_CB-HBT}(a),
where three Coulomb blockade peaks are plotted. For comparison, three
Coulomb blockade peaks are plotted for the AC-HBT case (Figure \ref{fig:current-biasing_effect_of_CB-HBT}(b)).
The CB-HBT amplified peaks are broadened by the current-biasing effect
and the blockade region never reaches zero current as it would with
a smaller constant voltage bias. The AC-HBT amplified peaks are much
narrower and minimally broaden due to having a constant, small voltage
bias regardless of HBT power. Comparable sensitivities can be achieved
for either circuit around 10 $\mu$A/V.

\section{Electron Temperature Measurement\label{sec:Electron_Temp_Meas}}

Heating of electrons in the quantum dot due to the operation of the
connected HBT is a concern, therefore we examined the dependence of
electron temperature on HBT amplifier bias (Figure \ref{fig:amp_comparison}(e)).
For the CB-HBT, The electron temperature of the QD was measured by
extracting the width of a Coulomb blockade peak as a function of fridge
temperature. The QD was tuned to a transport regime where the QD was
approximately equally tunnel-coupled to both reservoirs and there
were around 10 electrons in the QD. The source-drain bias was reduced
to 5 $\mu$V\textsubscript{rms} to avoid bias heating. A Coulomb
peak was chosen where a minimum width was observed in Coulomb diamond
measurements. After extracting the lever-arm of the gate used to measure
the broadening (13 $\mu$eV/mV), we find that the minimum linewidth
yields an electron temperature around 150 mK. Heating of the QD begins
where the CB-HBT is operating with over 100 gain, therefore the CB-HBT
circuit can amplify well while heating the electrons to 160--200
mK.

\begin{figure}
\begin{centering}
\includegraphics[width=8.5cm]{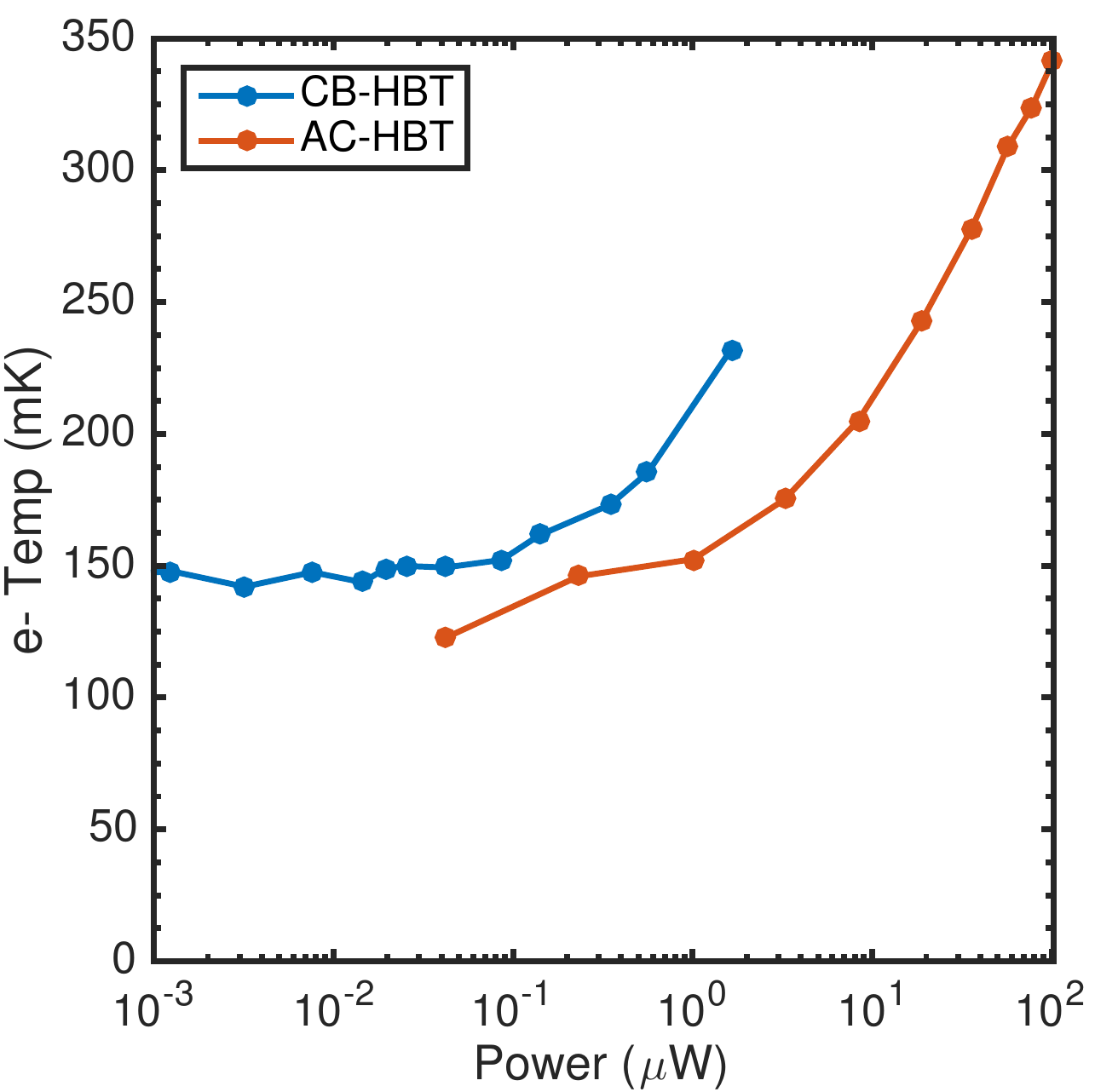}
\par\end{centering}
\caption{Electron temperature vs. power for both circuits. Base temperatures
are between 120--150 mK. Both circuits operate in the 160--200 mK
range for single-shot data taken.}

\label{fig:e_temp}
\end{figure}

For the AC-HBT setup, the base electron temperature was around 120
mK. This is confirmed by the measurements of the electron temperature
when measuring the SET signal directly through the shunt resistor
(R\textsubscript{S} in Figure \ref{fig:ac-hbt_circuit_diagram}(a))
with the HBT turned off. With the HBT on, The electron temperature
is deduced by measuring the Fermi-Dirac linewidth of the (1,0)-(2,0)
charge transition. When the AC-HBT bias is increased up to 3.24 $\mu$W,
the electron temperature remains near the base temperature (Figure
\ref{fig:e_temp}). For powers above this threshold, the electron
temperature increases approximately linearly with power. This might
be due to local heating of the PCB and wires, which increase the temperature
of the nearby device \citep{Knapp_2018}. No effort has been made
to heat sink the AC-HBT in this experiment, so further tests with
various heat sinking options will be performed to minimize the increase
in electron temperature. Nonetheless, an electron temperature of 200
mK is achieved for the bias condition that provides the minimum amplifier
noise.

\section{HBT Characterization\label{sec:HBT_Characterization}}

\begin{figure}
\begin{centering}
\includegraphics[width=8.5cm]{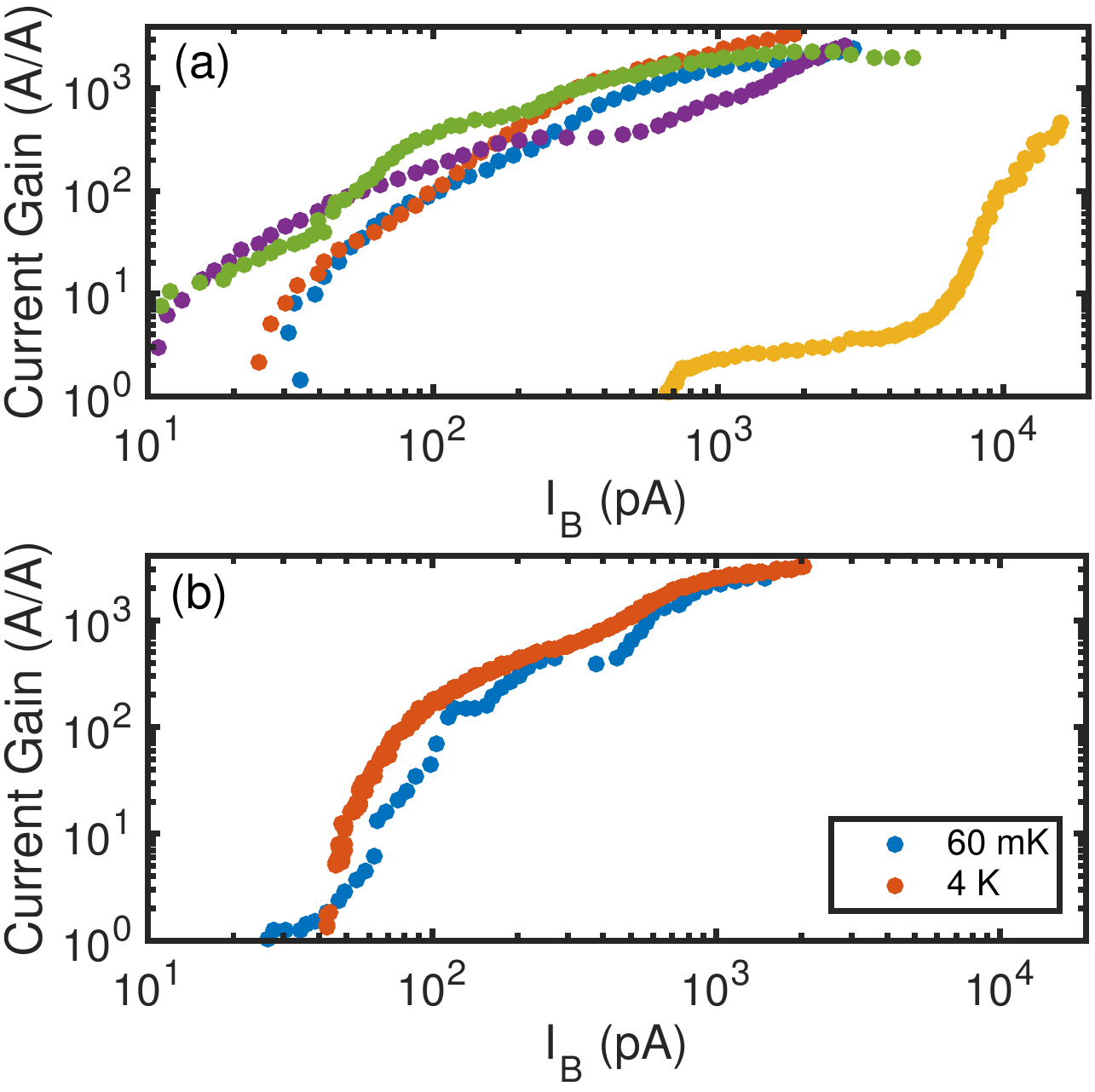}
\par\end{centering}
\caption{(a) Example plots of current gain vs. base current for different HBTs.
Several curves reach current gain > 1000 for base currents < 500 pA.
The HBT corresponding to the yellow curve is subpar since it requires
base current < 10 nA to reach current gain < 1000. (b) Current gain
vs. base current at different temperatures for HBT used in the CB-HBT
circuit. There is a slight difference in the two curves, however the
performance at 60 mK is enough to efficiently amplify and perform
single-shot readout.}

\label{fig:HBT_characterization}
\end{figure}

Before being used in either amplification circuit, HBTs are initially
characterized in liquid helium at 4 K using PCBs with eight HBTs mounted
on them. We find that HBT performance at 4 K---particularly current
gain vs. base current---changes minimally when HBTs are cooled down
to 20--60 mK in a dilution refrigerator (Figure \ref{fig:HBT_characterization}(b)).
This is most likely due to the charge-carrier transport mechanism
changing from a drift-diffusion regime (temperature dependent) to
a tunneling regime (barrier dependent) at around 30 K \citep{Davidovic_2017}.

In order to characterize HBTs, Keithley 2400 source-measure units
are used as current meters and connected to the HBT base and collector
terminals. A power supply (emitter bias) is connected to the HBT emitter
terminal and used to bias the HBT to different operating regimes.
The emitter bias has to reach approximately -1 V for the HBT to begin
operating in an amplifying regime. As the emitter bias is changed
from -1.00 V to around -1.07 V, the collector and base current begin
to increase exponentially. The current gain, defined by dividing the
collector current by the base current, also increases exponentially
as emitter bias changes.

Previous measurements without HBT amplification circuits indicate
that the SET current should be below several hundred pA in order to
avoid QD electron heating. For the CB-HBT, we select HBTs based on
their current gain at low base currents. Around 20\% of HBTs characterized
will have current gain > 100 at base current < 200 pA (Figure \ref{fig:HBT_characterization}(a)).
For the AC-HBT, the transconductance (g\textsubscript{m}) is the
only metric required for selection. Since the HBTs were fabricated
with g\textsubscript{m} as a primary metric, > 80\% of HBTs are usable
for the AC-HBT circuit even at low temperatures. However, g\textsubscript{m}
does not scale ideally in these HBTs at cryogenic temperatures. For
a given HBT, $g_{m}\propto I_{C}^{n}$, where $n=1$ in normal conditions.
In the HBTs used in this work, $n\approx0.8$, which leads to suboptimal
SNR at higher power.

\begin{figure}
\begin{centering}
\includegraphics[width=8.5cm]{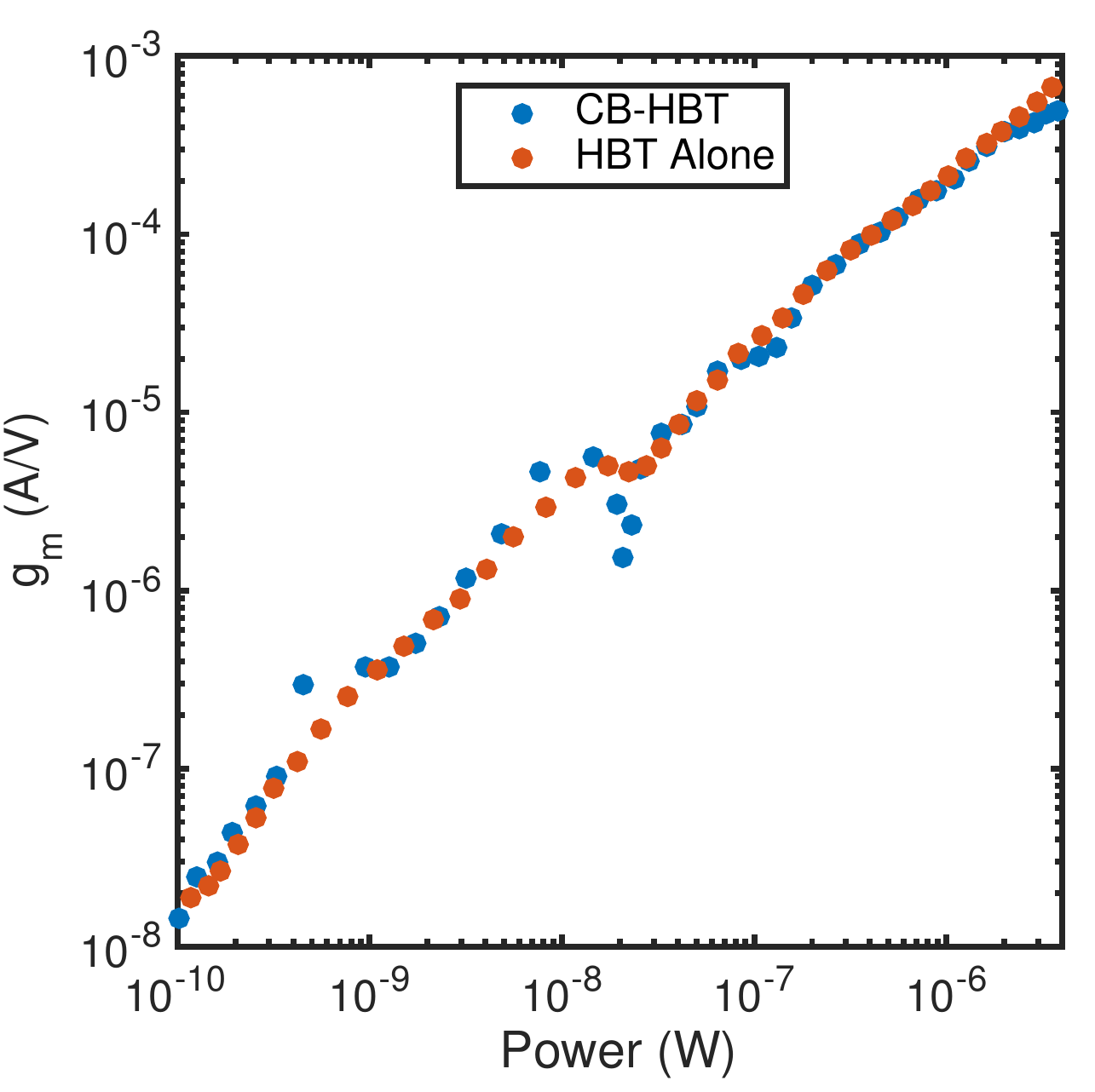}
\par\end{centering}
\caption{Transconductance vs. power for the CB-HBT (SET connected to HBT) and
the same HBT without an SET connected. The data overlaps for both
cases, therefore the transconductance can be reliably measured directly
in the CB-HBT (assuming $r_{set}\ll r_{\pi}$).}

\label{fig:gm_compare}
\end{figure}

\section{CB-HBT Small Signal Gain \label{sec:CB-HBT_Effective_Gain}}

The gain of the CB-HBT is calculated using a standard BJT small-signal
model. A small voltage fluctuation at the base node is usually converted
to a large current fluctuation at the collector node by the transconductance,
g\textsubscript{m} = $\frac{di_{c}}{dv_{be}}$. This voltage fluctuation
is usually the small-signal base-emitter junction resistance, r\textsubscript{\textgreek{p}},
multiplied by the base current. However, in the case of the CB-HBT,
$r_{set}\,||\,r_{\pi}$, therefore the parallel combination of the
two resistances is required to calculate gain:
\begin{equation}
\textrm{gai}\textrm{n}_{CB}=\frac{i_{c}}{i_{set}}=g_{m}(r_{set}\,||\,r_{\pi})\label{eq:cb-hbt_gain}
\end{equation}

\section{Noise Models \label{sec:Noise_Models}}

Sources of noise in the HBT amplification circuits include: shot noise,
Johnson noise, triboelectric noise associated with the coaxial lines
coupled to fridge vibration \citep{Kalra_2016}, room temperature
amplifier noise, and other instrumental noise. At relatively low power
operation regimes (< 1 $\mu$W for the AC-HBT and < 200 nW for the
CB-HBT), the noise due to vibrations in the fridge dominates at around
1 pA/$\sqrt{\textrm{Hz}}$. The input noise spectral density of the
room temperature amplifier is relatively low (100--500 fA/$\sqrt{\textrm{Hz}}$),
therefore we focus on noise sources much more dominant. When either
circuit is operating in a regime appropriate for single-shot readout,
the base shot noise is greater than the collector shot noise (Figures
\ref{fig:ac-hbt_circuit_diagram}(e) and \ref{fig:cb-hbt_circuit_diagram}(e)).
For the SET shot noise in either case, we do not consider a Fano factor,
which would reduce the noise for a given power \citep{Beenakker_2003,Kafanov_2009}.
The total noise for either circuit is calculated by assuming noise
sources are independent processes and adding noise sources in quadrature.

\begin{figure}
\begin{centering}
\includegraphics[width=8.5cm]{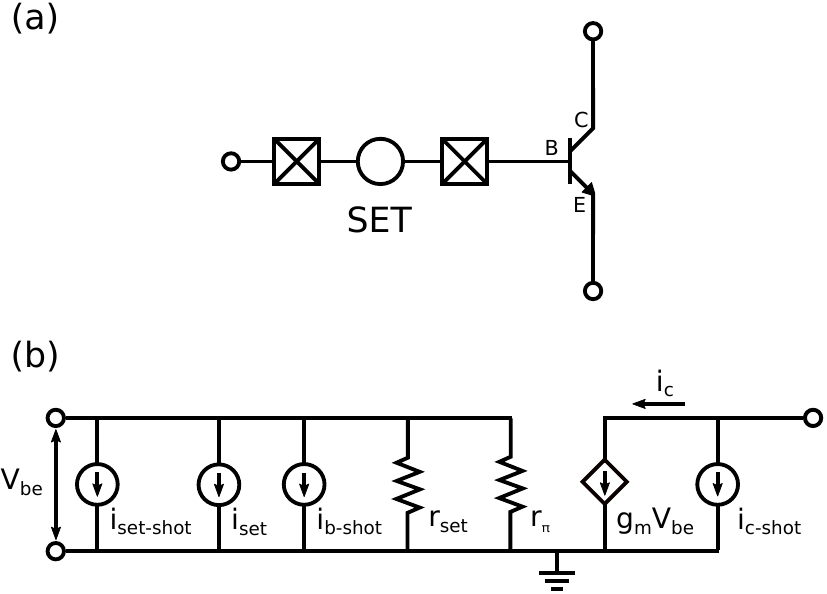}
\par\end{centering}
\caption{(a) CB-HBT circuit schematic for reference. (b) CB-HBT effective circuit
model. The shot noise current source, i\protect\textsubscript{b-shot},
is in parallel with r\protect\textsubscript{set} and r\protect\textsubscript{\textgreek{p}}.
Most of the shot noise does not enter the base of the HBT because
r\protect\textsubscript{set} <\textcompwordmark < r\protect\textsubscript{\textgreek{p}}.
The signal, i\protect\textsubscript{set}, is also shown, which is
amplified according to Equation \ref{eq:cb-hbt_gain}.}

\label{fig:cb-hbt_noise_model}
\end{figure}

Noise modeling for the CB-HBT circuit is nontrivial because of current
division at the HBT base node since $r_{set}\ll r_{\pi}$. The SET
and base current are reduced to a Norton equivalent circuit, and the
HBT is reduced to r\textsubscript{\textgreek{p}} connected to a current
source which takes voltage fluctuations (v\textsubscript{be}) across
r\textsubscript{\textgreek{p}} and converts them to collector current
via the transconductance, g\textsubscript{m}. For the CB-HBT, the
noise model is a shot noise current source ($i_{b-shot}=\sqrt{2\,e\,I_{B}\,\Delta f}$,
where $I_{B}$ is the DC base current, and $\Delta f$ is the bandwidth
centered on frequency $f$) in parallel with r\textsubscript{set}
and r\textsubscript{\textgreek{p}} (Figure \ref{fig:cb-hbt_noise_model}(b)).
Since $r_{set}\ll r_{\pi}$, most of the shot noise current goes through
the SET to ground, and a much smaller amount enters the HBT base and
is amplified. The amplified base shot noise is shown in Equation \ref{eq:cb-hbt_amplified_shot-noise}:
\begin{equation}
i_{\textrm{b-shot-amp}}=i_{\textrm{b-shot}}\:\textrm{gai}\textrm{n}_{CB}=i_{\textrm{b-shot}}\:g_{m}\:(r_{set}\,||\,r_{\pi})\label{eq:cb-hbt_amplified_shot-noise}
\end{equation}
 This amplified base shot noise is estimated in Figure \ref{fig:cb-hbt_circuit_diagram}(e)
as the orange curve where g\textsubscript{m} and r\textsubscript{\textgreek{p}}
are calculated from Gummel plots of the HBT and r\textsubscript{set}
is assumed to be 3 M\textgreek{W}, which was verified in later measurements
with the HBT disconnected from the Si-MOS device.

\begin{figure}
\begin{centering}
\includegraphics[width=8.5cm]{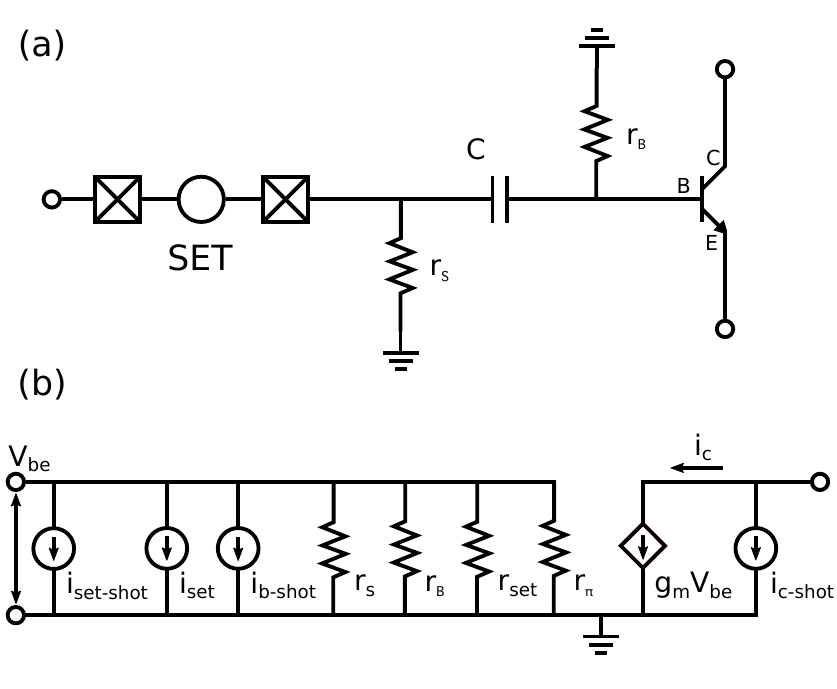}
\par\end{centering}
\caption{(a) AC-HBT circuit schematic for reference. (b) AC-HBT effective circuit
model with signal, i\protect\textsubscript{set}, also shown. The
model is similar to the CB-HBT with two new resistors added in parallel,
r\protect\textsubscript{S} and r\protect\textsubscript{B}.}

\label{fig:ac-hbt_noise_model}
\end{figure}

The noise model for the AC-HBT is similar to the CB-HBT with r\textsubscript{S}
and r\textsubscript{B} added in parallel to r\textsubscript{set}
and r\textsubscript{\textgreek{p}}. The coupling capacitor, C, is
considered a short at the frequencies appropriate to model noise in
the AC-HBT. The Johnson noise of R\textsubscript{S} in the AC-HBT
circuit is $v_{s-jn}=\sqrt{4\,k_{B}\,T\,R_{S}\,\Delta f}$ (where
$T$ is the temperature) and does not contribute significantly in
the single-shot operation regime. Since the AC-HBT has a separate
current to bias the base-emitter junction, $I_{SET}\neq I_{B}$, therefore
the base shot noise and SET shot noise are considered separately.
However, $I_{SET}<I_{B}$, so the base shot noise is always dominant
in amplifying regimes.

\section{AC-HBT Bias Tee Parameters \label{sec:AC-HBT-Bias-Tee}}

The bias tee parameters for the AC-HBT were chosen to be R\textsubscript{S}
= 100 k\textgreek{W} and C = 10 nF, which sets a high pass filter
at 160 Hz. Operating the circuit at frequencies higher than 160 Hz
aids in avoiding higher noise levels at lower frequency due to 1/f-like
noise behavior in the system.

The shunt resistance value is chosen to be less than r\textsubscript{set}
(100s of k\textgreek{W}) so that most of the SET bias voltage drops
across the SET.
\end{document}